\documentclass[12pt]{article}
\usepackage{axodraw,amsmath,amssymb,exscale,epsfig}

\begin{document}
\parskip 0.3cm

\def\simgt{\mathrel{\lower2.5pt\vbox{\lineskip=0pt\baselineskip=0pt
           \hbox{$>$}\hbox{$\sim$}}}}
\def\simlt{\mathrel{\lower2.5pt\vbox{\lineskip=0pt\baselineskip=0pt
           \hbox{$<$}\hbox{$\sim$}}}}
\newcommand{\Tr}{\mathop{\rm Tr}}   
\def\dsl{\hbox{/\kern-.6800em$D$}}
\def\rsl{\hbox{/\kern-.6800em$R$}}


\begin{titlepage}

\begin{flushright}
FT-UAM-03-11 \\
FERMILAB-Pub-03/195-T \\
UAB-FT-549
\end{flushright}

\vskip 1.3cm

\begin{center}

{\Large \bf Higgs as a Holographic Pseudo-Goldstone Boson}

\vskip 0.7cm

{\large Roberto Contino$^a$, Yasunori Nomura$^b$ and Alex Pomarol$^c$}

\vskip 0.4cm

$^a$ {\it Departamento de F{\'\i}sica Te{\'o}rica C-XI, 
          Universidad Aut{\'o}noma de Madrid,\\ 
          Cantoblanco, 28049 Madrid, Spain} \\
$^b$ {\it Theoretical Physics Department, Fermi National Accelerator 
          Laboratory, \\ Batavia, IL 60510, USA} \\
$^c$ {\it IFAE, Universitat Aut{\`o}noma de Barcelona, 
          08193 Bellaterra (Barcelona), Spain}

\vskip 1.5cm

\abstract{
The AdS/CFT correspondence allows to relate 4D strongly coupled 
theories to weakly coupled theories in 5D AdS. We use this 
correspondence to study a scenario in which the Higgs appears 
as a composite pseudo-Goldstone boson (PGB) of a strongly coupled 
theory. We show how a non-linearly realized global symmetry protects 
the Higgs mass and guarantees the absence of quadratic divergences 
at any loop order. The gauge and Yukawa interactions for the PGB 
Higgs are introduced in a simply way in the 5D AdS theory, and 
their one-loop contributions to the Higgs potential are calculated 
using perturbation theory. These contributions are finite, giving 
a squared-mass to the Higgs which is one-loop smaller than the mass 
of the first Kaluza-Klein state. We also show that if the symmetry 
breaking is caused by boundary conditions in the extra dimension, 
the PGB Higgs corresponds to the fifth component of the bulk gauge 
boson. To make the model fully realistic, a tree-level Higgs quartic 
coupling must be induced. We present a possible mechanism to generate 
it and discuss the conditions under which an unwanted large Higgs 
mass term is avoided.}

\end{center}
\end{titlepage}

\section{Introduction}

Despite its tremendous phenomenological success, the standard 
model is almost certainly not a fundamental theory of nature.
Quantum instabilities of the Higgs potential strongly suggest that 
it must be replaced by some other theory at energies not much higher 
than the electroweak scale. Such a theory, for example, can 
be a supersymmetric theory~\cite{Martin:1997ns}, a strongly 
coupled gauge theory~\cite{Weinberg:gm}, or a theory of quantum 
gravity~\cite{Arkani-Hamed:1998rs}. In these theories the quadratic 
divergence of the Higgs squared-mass parameter is cut off either 
by embedding the Higgs boson into some larger multiplet or by 
giving it an internal structure. A physical scale then exists, 
$\Lambda_{\rm NP}$, at which many new particles appear revealing 
the underlying symmetry or dynamics.

In this paper we consider a class of theories where the standard-model 
Higgs boson arises as a composite particle of some strongly coupled 
dynamics. The dynamical scale $\Lambda_{\rm NP}$ then must be 
parametrically larger than the scale of the Higgs mass in order 
to avoid strong constraints coming from direct and indirect searches 
of new particles at colliders. This suggests that the Higgs mass must 
be protected by some (approximate) symmetry even below the scale 
$\Lambda_{\rm NP}$. A natural candidate for such a symmetry is an 
internal global symmetry under which the Higgs boson transforms 
non-linearly: the Higgs is a pseudo-Goldstone boson (PGB) of the 
broken global symmetry.  This situation is somewhat similar to 
that of the pion in QCD, although for the Higgs there are additional 
requirements. It must have a sizable quartic self-coupling and 
appropriate Yukawa couplings to the quarks and leptons. In this paper 
we aim to build theories of this kind, which are well under control 
as effective field theories, and in which some quantities are even 
calculable despite the strongly interacting dynamics.

The basic observation is the following.  Suppose we have a strongly 
coupled gauge theory that produces the Higgs boson as a composite 
state. In general, it is quite difficult to obtain quantitative 
low-energy predictions in such a theory because of non-perturbative 
effects; one can at best derive estimates by using certain scaling 
arguments. This is indeed the case if the gauge interaction is 
asymptotically free, as in QCD. However, it is not necessarily true 
if the theory remains strongly coupled in the UV and approaches 
a non-trivial conformal fixed point. In this case it is possible 
that the theory, in the limit of large number of ``colors'', 
has an equivalent description in terms of a weakly coupled 
5D theory defined on the truncated anti de-Sitter (AdS) 
space~\cite{Maldacena:1997re,Arkani-Hamed:2000ds}. This allows us 
to construct theories where the Higgs boson is interpreted as a 
composite state of a strong dynamics and yet some physical quantities 
such as the Higgs potential can be computed using perturbation theory.

The actual implementation of the above idea is quite simple, as far 
as gauge and Yukawa interactions are concerned.  These interactions 
explicitly break the global symmetry, but, as we will see, they do 
not induce quadratic divergences for the Higgs mass at any loop order. 
The global symmetry protects the Higgs mass parameter at high energies 
and predicts it to be a one-loop factor smaller than the mass of the 
lightest resonance $\sim \Lambda_{\rm NP}$. This property makes this 
class of theories quite interesting.

To make the model fully realistic, however, we must generate 
a sizable Higgs quartic coupling. This is crucial not only 
to obtain a large enough physical Higgs mass, but also to 
obtain the needed mas gap between the electroweak scale and 
$\Lambda_{\rm NP}$ (as suggested by experiments). We will propose 
a mechanism that allows to generate a Higgs quartic coupling at 
tree level. This mechanism needs specific assumptions on the form 
of the explicit breaking of the global symmetry. However, these 
are assumptions on the underlying physics around the Planck scale,
and not on the TeV-scale physics which yields the Higgs boson as 
a composite particle. Therefore, once the particular pattern of 
breaking is assumed (which is not quite unnatural from the viewpoint 
of the 4D picture), we can compute the Higgs potential generated 
at loop level through the explicit symmetry-breaking effects.

Since the cutoff of our theory is around the Planck scale, there 
is no obstacle in extending the theory beyond $\Lambda_{\rm NP}$, 
up to very high energies. In this respect, our framework may be 
viewed as a way to provide a UV completion for ``little Higgs'' 
theories~\cite{Arkani-Hamed:2001nc} in which realistic models 
of the PGB Higgs have been constructed. It might be interesting 
to construct a UV completion of the existing little Higgs 
models~\cite{Arkani-Hamed:2002qy} using a warped spacetime 
as outlined in the present paper.

In the next section we start by defining the framework in more detail. 
We describe the basic structure of our theories in terms of both 4D 
and 5D pictures.  An explicit model is given in section~\ref{sec:higgsonb}, 
in which the Higgs boson is identified with a PGB arising from a scalar 
field located on the infrared brane. We present a possible mechanism 
to obtain a sizable Higgs quartic coupling and discuss the size of 
quantum corrections to the Higgs potential, which trigger electroweak 
symmetry breaking. In section~\ref{sec:A5higgs} we consider theories 
where the Higgs boson arises from the extra-dimensional component 
of a gauge field in a warped 5D spacetime. We point out that also in 
this case the Higgs is interpreted as a composite PGB in the 4D picture, 
and we present several realistic models. Conclusions are given in 
section~\ref{sec:concl}.

\section{Higgs as a Composite Pseudo-Goldstone Boson}

In this section we describe a class of theories where the standard-model 
Higgs arises as a composite PGB of a strong interaction, and in which 
the presence of a weakly coupled dual description allows the computation 
of certain quantities. We begin with the 4D description of our theory, 
which we also refer to as the holographic theory.  In this picture the 
theory consists of two sectors. One is a sector of elementary particles 
that correspond to the standard-model gauge bosons and (some of the) 
quarks and leptons. The other is a strongly coupled conformal field 
theory (CFT) sector, where the conformal symmetry is broken at low 
energies $1/L_1 \ll M_{\rm Pl}$. This sector produces CFT bound states 
due to the strong dynamics at the scale $1/L_1$. The Higgs will be one 
of these bound states.

To have a little hierarchy between $1/L_1$ and the Higgs mass, we 
require the Higgs to be a Goldstone boson of the CFT sector. For this 
purpose, we assume that the CFT has a global symmetry larger than 
the standard-model electroweak gauge group SU(2)$_L \times$U(1)$_Y$.
We find that it must be at least SU(3). If a global SU(3)$_L$ symmetry 
is spontaneously broken to SU(2)$_L$ by the CFT strong dynamics, 
then 5 Goldstone bosons appear, a doublet and a singlet under 
SU(2)$_L$. The doublet will be associated with the Higgs boson. 
The SU(3)$_L$ is not a symmetry of the standard-model gauge and 
matter fields which belong to the elementary sector, so that the 
couplings of these fields with the CFT explicitly break the global 
SU(3)$_L$ invariance. A mass for the Higgs boson is generated at 
loop level, which, if negative, will trigger electroweak symmetry 
breaking. The loop factor appearing in the Higgs mass can give 
a rationale for the little hierarchy between the electroweak scale 
and the compositeness scale $1/L_1$, although to perform quantitative 
computations we must go to the weakly coupled dual description 
of the theory.

The AdS/CFT correspondence~\cite{Maldacena:1997re}, as applied to 
a spontaneously broken CFT with a UV cutoff~\cite{Arkani-Hamed:2000ds}, 
allows us to relate the above scenario to a theory in a slice of 5D AdS. 
In this AdS picture the theory is weakly coupled and we can perform 
explicit calculations. The metric of the spacetime is given 
by~\cite{Randall:1999ee}
\begin{equation}
  ds^2 = \frac{1}{(kz)^2} \left(\eta_{\mu\nu}\, dx^\mu dx^\nu - dz^2\right)
  \equiv g_{MN}\, dx^M dx^N.
\label{eq:metric}
\end{equation}
The 5D coordinates are labeled by capital Latin letters, $M=(\mu,5)$ 
where $\mu=0,\dots,3$; $z = x^5$ represents the coordinate 
for the fifth dimension and $1/k$ is the AdS curvature radius. This 
spacetime has two boundaries at $z=L_0 \equiv 1/k \sim 1/M_{\rm Pl}$ 
(Planck brane) and $z = L_1 \sim 1/{\rm TeV}$ (TeV brane): the theory 
is defined on the line segment $L_0 \leq z \leq L_1$. 

The global symmetry of the 4D CFT is realized as a bulk gauge symmetry 
in the 5D picture. In the case of a 4D theory where the CFT sector has 
a global SU(3)$_L$ invariance, the dual theory is a 5D SU(3)$_L$ gauge 
theory. This SU(3)$_L$ is spontaneously broken by two scalars. 
One living on the Planck brane and the other on the TeV brane. 
Being separated in space, these scalars do not ``see'' each other 
at the classical level, so that the theory contains, in the gaugeless 
limit, an enlarged global SU(3)$\times$SU(3) symmetry. By giving vacuum 
expectation values (VEVs) to the two scalars, the global symmetry is 
spontaneously broken, SU(3)$\times$SU(3) $\rightarrow$ SU(2)$\times$SU(2), 
delivering two sets of 5 Goldstone bosons. When we gauge the SU(3)$_L$ 
subgroup of SU(3)$\times$SU(3), the Goldstone bosons which parametrize 
the SU(3)$_L/$SU(2)$_L$ space are true Goldstone bosons and are eaten 
to form massive gauge vectors. The remaining ones are PGBs, since 
gauge interactions do not respect the full global SU(3)$\times$SU(3).
In a slice of warped space the scalar living on the Planck brane 
corresponds, to a very good approximation, to the true Goldstone 
boson.  This is because its VEV will be naturally of order $M_{\rm Pl}$, 
much larger than the VEV of the scalar on the TeV brane.  Therefore, 
the scalar on the TeV brane corresponds to the PGB, which we identify 
as the standard-model Higgs boson.

The holographic dual of this 5D setup is thus the 4D theory we 
described before: the Higgs corresponds to the composite Goldstone 
boson of a CFT sector whose global SU(3)$_L$ invariance is 
spontaneously broken down to SU(2)$_L$ by the strong dynamics.
An explicit breaking of the global CFT invariance is communicated 
by the interactions with the elementary sector, and a mass for the 
Goldstone bosons is generated at one loop. The spontaneous symmetry 
breaking of the CFT sector corresponds to the TeV-brane breaking of 
the 5D theory, while the explicit breaking given by the elementary 
sector is associated with the Planck-brane dynamics. This means that 
any process of the holographic theory where the explicit breaking is 
communicated from the elementary sector to the CFT, will correspond 
in the AdS picture to some kind of transmission from the Planck brane 
to the TeV brane. The mass of the PGB is an important example: 
non-locality in the 5D theory implies that it is a calculable and 
finite effect. This is a crucial ingredient of our class of theories, 
which gives a rationale for explaining the little hierarchy. We will 
come back to this point in the next section, where we compute the 
Higgs mass at one loop.

The scalar on the Planck brane can be replaced by boundary conditions 
that break SU(3)$_L$ to SU(2)$_L$ on the Planck brane.  The breaking 
on the TeV brane can also be realized by boundary conditions. In this 
case, the PGB corresponds to the fifth component of the gauge boson 
as we will see in section~\ref{sec:A5higgs}.

A realistic theory must have Yukawa couplings between the Higgs and 
quarks and leptons. It is simple to incorporate this feature in our 
theories.  The fermions must be put in the bulk (at least one of 
their chiralities) in representations of SU(3)$_L$ and be coupled 
to the Higgs on the TeV brane. After the SU(3)$_L$ breaking on the 
Planck brane, we can obtain the standard-model quarks and leptons 
as the only massless fermions (before electroweak symmetry breaking). 
Large enough Yukawa couplings are obtained if the bulk fermion masses 
are in a certain range such that they probe the SU(3)$_L$ breaking 
effect on the Planck brane. These Yukawa couplings then induce 
a negative one-loop contribution to the Higgs mass term, which can 
trigger electroweak symmetry breaking.

Models with the Higgs as a PGB face, however, a significant 
phenomenological challenge.  The Higgs quartic coupling $\lambda_H$ 
is induced only at one-loop level, giving a physical Higgs mass, 
$m_h^2=2\lambda_H \langle H\rangle^2$, below the experimental bound 
of $m_h \simgt 114~{\rm GeV}$. This is one of the major obstacles for 
the realization of the Higgs as a PGB. A realistic model has to induce 
$\lambda_H$ at tree level. The challenge is to induce this coupling 
without generating a tree-level mass term; otherwise the little 
hierarchy between the electroweak scale and the compositeness scale 
is lost. Below we present a possible mechanism, although, as we will 
explain, it requires further assumptions about the symmetry breaking 
physics at high energies to keep the PGB massless at tree level.

\section{A Model}
\label{sec:higgsonb} 

In this section we present an explicit model that leads to the 
standard model with the Higgs boson as a PGB.  This represents
a concrete example for the general theories introduced in the 
previous section. We discuss an alternative possibility in the 
next section, where the Higgs is identified with the extra-dimensional 
component of the gauge boson, $A_5$.

We consider a 5D theory in a slice of AdS with a gauge symmetry 
SU(3)$_L \times$U(1)$_X$, which contains the electroweak
SU(2)$_L \times$U(1)$_Y$ as a subgroup.  The extra U(1)$_X$ is 
introduced to give the correct hypercharges to the fermions. 
All the gauge bosons are assumed to have Neumann boundary conditions 
at both branes (we will work in the unitary gauge $A_5=0$). 
The matter content of the model is the following (we only consider 
the third-generation quark sector for simplicity, but the extension 
to the full standard model is straightforward).  We introduce two 
bulk fermions $Q$ and $D$ which transform as ${\bf 3^*_{1/3}}$ and 
${\bf 3_0}$ under SU(3)$_L \times$U(1)$_X$. Since the bulk fermions 
$\Psi$ are in the Dirac representation, they are decomposed into the 
left-handed, $\Psi_L$, and right-handed, $\Psi_R$, components in terms 
of the 4D irreducible representation (Weyl fermion): $Q = Q_L + Q_R$ 
and $D = D_L + D_R$.  The boundary conditions for these fields are 
chosen such that $Q_L$ and $D_R$ ($Q_R$ and $D_L$) obey Neumann 
(Dirichlet) boundary conditions at the Planck and TeV branes.  
This implies that only $Q_L$ and $D_R$ have massless zero modes.  
We also introduce the boundary fermion $U_R$ on the TeV brane, which 
transforms as ${\bf 1_{2/3}}$ under SU(3)$_L \times$U(1)$_X$.

We assume that SU(3)$_L \times$U(1)$_X$ is broken on the Planck 
brane to the standard-model group SU(2)$_L \times$U(1)$_Y$, with 
$Y=T_8/\sqrt{3}+X$.  This breaking can be caused, for example, 
by a scalar $S$ on the Planck brane with quantum numbers 
${\bf 3_{1/3}}$ under SU(3)$_L \times$U(1)$_X$. The bulk fermion 
fields are decomposed under the standard-model group as
\begin{equation*} 
\begin{split}
  Q ({\bf 3^*_{1/3}}) 
  &= Q_L^{(D)}({\bf 2^*_{1/6}}) + Q_L^{(S)}({\bf 1_{2/3}})
    + Q_R^{(D)}({\bf 2^*_{1/6}}) + Q_R^{(S)}({\bf 1_{2/3}}) ,
\\[0.2cm]
  D({\bf 3_0})
  &= D_L^{(D)}({\bf 2_{1/6}}) + D_L^{(S)}({\bf 1_{-1/3}})
    + D_R^{(D)}({\bf 2_{1/6}}) + D_R^{(S)}({\bf 1_{-1/3}}) .
\end{split}
\end{equation*}
We assume that the symmetry breaking dynamics at the Planck brane 
is such that only $Q_L^{(D)}$ and $D_R^{(S)}$ are left as 4D massless 
fields, and we identify these fields with the standard-model doublet 
quark $q_L$ and singlet bottom quark $b_R$.\footnote{
This situation can be realized, for example, by adding the following 
couplings on the Planck brane: $S^\dagger \bar{Q}_L U^\prime_R 
+ M \bar{D}_R D^\prime_L + S \bar{D}^\prime_L D^{\prime\prime}_R$, 
where $U^\prime_R({\bf 1_{2/3}})$, $D^\prime_L({\bf 3_{0}})$ and 
$D^{\prime\prime}_R({\bf 1_{-1/3}})$ are extra quarks that marry with 
the unwanted SU(3)$_L$ partners of the zero modes of $Q_L^{(D)}$ and 
$D_R^{(S)}$. This symmetry breaking pattern can be more easily 
obtained if we choose to break SU(3)$_L$ through boundary conditions. 
The details of this breaking, however, are not relevant for physics 
at the TeV scale.}
The standard-model singlet top quark, $t_R$, is identified with the 
brane field $U_R$. When the Yukawa couplings on the TeV brane are 
included, however, $q_L$ will be a mixture of $Q_L^{(D)}$ and 
$D_L^{(D)}$ states, while $t_R$ will be a mixture of $U_R$ and 
$Q_R^{(S)}$ states, as we will see below. The color SU(3)$_C$ 
can be introduced in a straightforward way: $Q$, $D$ and $U$ 
all transform as ${\bf 3}$ of SU(3)$_C$.

What about the Higgs field?  To have the Higgs as a PGB, we introduce 
a scalar field $\Sigma$ on the TeV brane transforming as ${\bf 3_{1/3}}$ 
under SU(3)$_L \times$U(1)$_X$. We assume that $\Sigma$ gets a VEV, 
say by the potential ${\cal L} = \delta(z-L_1) \sqrt{-g_{\rm ind}} 
(\Sigma^\dagger \Sigma - \tilde{v}^2)^2$ where $g_{\rm ind}$ is the 
induced metric on the brane. We can then parametrize the scalar field 
$\Sigma$ as
\begin{equation}
  \Sigma ({\bf 3_{1/3}}) 
  = e^{iT^aG^a} 
  \left(\begin{array}{c} 0 \\ 0 \\ \tilde{v}+\varphi \end{array}\right),
  \qquad T^aG^a=\frac{1}{\tilde{v}}
  \left(\begin{array}{cc} 0 & H \\ H^\dagger & \eta \end{array}\right),
\label{eq:sigmadef}
\end{equation}
where the fields $H$ and $\eta$ are PGBs that transform respectively 
as doublet and singlet of SU(2)$_L$, and $\varphi$ is a real scalar 
field; $\tilde{v}$ is the VEV of $\Sigma$ in terms of the 5D metric. 
We also define, for later convenience, the VEV in terms of the 4D 
metric $v \equiv \tilde{v} L_0/L_1$, which takes a value of the order 
of the local cutoff on the TeV brane, $\Lambda_{\rm IR} \sim {\rm TeV}$.
The unbroken group under the $\Sigma$ VEV can naturally be aligned 
with that under the $S$ VEV, due to possible radiative effects 
relating them. We then find that the PGB field $H$ has the 
appropriate SU(2)$_L \times$U(1)$_Y$ quantum numbers, ${\bf 2_{1/2}}$, 
to be identified as the standard-model Higgs boson.

We now proceed to the interaction terms of the theory. The 5D theory 
has mass terms for the bulk fermions:
\begin{equation}
  {\cal L} = \sqrt{g}\, \left[{\cal L}_{\rm kin}
  - M_Q\, \bar{Q}_L Q_R - M_D\, \bar{D}_L D_R\right],
\end{equation}
where ${\cal L}_{\rm kin}$ are the kinetic terms (see Eq.~(\ref{eq:Lfermion}) 
for the explicit expression). These mass terms control the shape of 
the wavefunctions of the $Q_L$ and $D_R$ zero modes, and hence the size 
of the various low-energy 4D couplings. In addition, we introduce the 
Yukawa couplings for the matter fields on the TeV brane:
\begin{equation}
  {\cal L} = \delta(z-L_1) \sqrt{-g_{\rm ind}} 
  \left[ \lambda_U\, \Sigma^\dagger \bar{Q}_L U_R 
  + \lambda_D\, \Sigma^i \bar{Q}_L^j D_R^k \epsilon^{ijk} 
  + {\rm h.c.} \right],
\end{equation}
where $i,j,k$ represent the SU(3)$_L$ index. After integrating out the 
Kaluza-Klein (KK) states, we obtain the following effective Lagrangian: 
\begin{equation}
\begin{split}
  {\cal L}_{\rm 4D} 
  =& Z_H |D_\mu H|^2 + iZ_Q\,\bar{q}_L\!\not\!\partial\, q_L 
    + i\bar{b}_R\!\not\!\partial\, b_R 
    + iZ_U\, \bar{t}_R\!\not\!\partial\, t_R 
\\
   & -i\lambda_U f_{Q}\, H^\dagger \bar{q}_L t_R 
    + i\lambda_D f_Q f_D\, H \bar{q}_L b_R + {\rm h.c.}
\end{split}
\end{equation}
Here, $f_{Q}$ ($f_D$) denotes the value of the zero-mode wavefunction 
of $Q_L^{(D)}$ ($D_R^{(S)}$) at the TeV brane; one has
\begin{equation}
  f_i = \frac{1}{N_0}\left(\frac{L_1}{L_0}\right)^{(1/2\mp c_i)}, \ \ \ \ 
  N_0^2 = \frac{L_0}{(1\mp 2c_i)}
    \left[\left(\frac{L_1}{L_0}\right)^{(1\mp 2c_i)}-1\right],
\label{eq:wfun}
\end{equation}
where the $\mp$ sign holds for the zero mode of a left (right) handed 
field and $c_i=M_i/k$ for $i=Q,D$. The factors $Z_H$, $Z_Q$ and $Z_U$ 
arise respectively due to the mixing of $H$ with the heavy KK gauge 
bosons, the mixing of the zero mode of $Q^{(D)}_L$ with the KK 
states of $D^{(D)}_R$, and the mixing of $U_R$ with the KK states of 
$Q_L^{(S)}$. These mixings appear when $\Sigma$ gets a VEV, and we find
\begin{equation}
  Z_H^{-1} = 1 + \frac{(g_5 \tilde{v})^2}{4k},
\ \ \
  Z_Q =1+|f_Q\lambda_D\, \tilde{v}|^2 G_{D^{(D)}_R},
\ \ \
  Z_U =1+|\lambda_U\, \tilde{v}|^2 G_{Q^{(S)}_L}.
\end{equation}
Here, $G_{Q^{(S)}_L}$ ($G_{D^{(D)}_R}$) is the 5D propagator of 
$Q_L^{(S)}$ ($D_R^{(D)}$) evaluated on the TeV brane at zero 4D 
momentum: $G_{Q^{(S)}_L}=\hat{G}_{L}^{(-,+)}(L_1,L_1;0)$ and 
$G_{D^{(D)}_R}=\hat{G}_{R}^{(-,+)}(L_1,L_1;0)$, where the 
propagators $\hat{G}_{L,R}^{(-,+)}(z,z^\prime;p)$ can be found in 
the Appendix. After canonically normalizing the fields we obtain 
the top and bottom Yukawa couplings
\begin{equation}
  h_t = - \frac{i\lambda_U f_{Q}}{\sqrt{Z_H Z_QZ_U}},
\qquad
  h_b = \frac{i\lambda_Df_Qf_D}{\sqrt{Z_HZ_Q}}.
\label{eq:4d-yukawa}
\end{equation}
The Yukawa couplings strongly depend on $M_Q$ and $M_D$. For example, 
the dependence of the top Yukawa coupling on $M_Q$ is given by
\begin{equation} 
  h_t \sim \begin{cases} 
    \left(L_0/L_1\right)^{|c_Q|-1/2} & \text{for} \quad |c_Q|>1/2, \\[0.1cm]
    [\log(L_1/L_0)]^{-1/2}           & \text{for} \quad |c_Q|=1/2, \\[0.1cm]
    \mathcal{O}(1)                   & \text{for} \quad |c_Q|<1/2,
  \end{cases}
\label{eq:ht}
\end{equation}
and therefore the theory can be realistic only if $|M_Q| \simlt k/2$. 
It can also be shown that $|M_D| \simlt k/2$ is also needed to obtain 
realistic Yukawa couplings.

The holographic picture offers a simple explanation for the behavior 
of Eq.~(\ref{eq:ht}). For $c_Q>1/2$ the holographic theory consists 
of a CFT sector coupled to a left-handed dynamical ``source'' 
$\chi_L$ which transforms as a doublet of SU(2)$_L$~\cite{toappear}:
\begin{equation}
  {\cal L} = {\cal L}_\text{CFT} + i \bar{\chi}_L \!\not\!\partial\, \chi_L 
    + \lambda\, k^{1/2-c_Q}\, \bar{\chi}_L \cdot {\cal O}_R 
    + {\rm h.c.} + \cdots.
\label{eq:holo1}
\end{equation}
Here, $\lambda$ is a dimensionless coupling, and the ellipses stand 
for higher order operators ($M_{\rm Pl}$-suppressed), and gauge and 
bottom interactions. The coupling $\bar\chi_L \cdot {\cal O}_R$ induces 
a tree-level mixing between the elementary source and the CFT bound 
states; its strength is determined by the AdS/CFT correspondence, 
which relates the dimension of ${\cal O}_R$ with the value of $c_Q$: 
${\rm dim}[{\cal O}_R]=3/2+|c_Q+1/2|$. For $c_Q>1/2$ the coupling is 
always irrelevant, so that the physical massless eigenstate $q_L$, 
to be identified with the standard-model quark doublet, is almost 
the elementary state $\chi_L$. The quark singlet $t_R$ appears in 
the theory as a composite CFT state. For $c_Q<-1/2$ the holographic 
picture is very different. We find that the 4D theory must contain 
a right-handed dynamical source $\chi_R$ which is a singlet of 
SU(2)$_L$~\cite{toappear}:
\begin{equation}
  {\cal L} = {\cal L}_\text{CFT} 
    + i \bar \chi_R \!\not\!\partial\, \chi_R 
    + \lambda\, k^{1/2+c_Q}\, \bar \chi_R\cdot {\cal O}_L 
    + {\rm h.c.} + \cdots.
\label{eq:holo2}
\end{equation}
The AdS/CFT correspondence requires ${\rm dim}[{\cal O}_L]=3/2+|c_Q-1/2|$,
and the coupling between the elementary source and the CFT is again
irrelevant for $c_Q<-1/2$. The physical $q_L$ appears now as a composite 
CFT state, while $t_R$ is almost the elementary state $\chi_R$.

How does the Yukawa coupling between the composite Higgs $H$ and 
the physical quarks $q_L$ and $t_R$ arise?  As is clear from 
Eqs.~(\ref{eq:holo1}) and (\ref{eq:holo2}), the global SU(3)$_L$ 
invariance of the CFT is not a symmetry of the elementary sector, 
and the explicit breaking is communicated to the conformal sector 
through its coupling with the source $\chi_{L}$ or $\chi_{R}$.
The Yukawa coupling is then generated only through the insertion 
of the composite operators ${\cal O}_{L,R}$, since the process must 
involve the elementary source. This implies that the top Yukawa 
coupling must be proportional to $\lambda\, k^{1/2-|c_Q|}$, and 
therefore $h_t \sim \lambda\, (L_1/L_0)^{1/2-|c_Q|}$ as obtained 
in the 5D picture, Eq.~(\ref{eq:ht}).

For $-1/2 \leq c_Q \leq 1/2$ either of the two descriptions, 
Eq.~(\ref{eq:holo1}) or Eq.~(\ref{eq:holo2}), is valid. In this case 
the coupling of the CFT to the elementary sector is relevant (marginal 
for $c_Q=\pm 1/2$). This implies that the both physical quarks $q_L$ 
and $t_R$ are almost composite states. The only way to generate the 
top Yukawa is then through the virtual exchange of the source. 
The source has now a relevant coupling with the CFT, so that 
the CFT correction to the elementary propagator becomes important.
By resumming the infinite series of CFT insertions, one can express 
this correction as a renormalization of the coupling $\lambda$; 
for example, in the holographic picture of Eq.~(\ref{eq:holo1}), 
the conformal invariance gives
\begin{equation}
  \lambda^2(\mu) \sim 
  \frac{\lambda^2(k)}{1+\lambda^2(k) \left(\mu^2/k^2\right)^{c_Q-1/2}}.
\end{equation} 
Therefore, for $|c_Q| < 1/2$ and $\mu \ll k$, one has $\lambda(\mu) 
\sim (\mu/k)^{1/2-|c_Q|}$, so that the top Yukawa coupling,
proportional to $\lambda\, k^{1/2-|c_Q|}$, is always of order one. 
The particular case $|c_Q|=1/2$ gives a logarithmic suppression 
$h_t \propto [\ln(L_1/L_0)]^{-1/2}$.

Summarizing, we have obtained an SU(3)$_C \times$SU(2)$_L \times$U(1)$_Y$ 
gauge theory with a massless Higgs boson, $H$, a massless quark doublet, 
$q_L$, and two massless singlet quarks, $t_R$ and $b_R$, which have 
the Yukawa couplings of Eq.~(\ref{eq:4d-yukawa}).\footnote{
In the model presented here, there is also an extra singlet PGB 
$\eta$, which obtains a mass at one-loop level.}
The fermion content thus reproduces the third generation quark sector 
of the standard model. To complete the standard-model structure, however, 
we have to discuss the Higgs potential.  We address this remaining issue 
in the next two subsections.

\subsection{One-loop contribution to the Higgs potential}

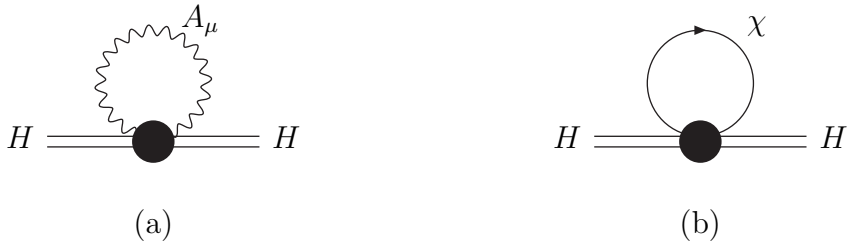
\begin{figure}[t]
\begin{minipage}[t]{0.47\linewidth}
  \begin{center}
  \begin{picture}(20,60)(0,0)
    \Line(-30,22)(50,22)
    \Line(-30,18)(50,18)
    \Text(-35,22)[r]{$H$}
    \Text(55,22)[l]{$H$}
    \Text(35,60)[br]{$A_\mu$}
    \PhotonArc(10,42)(20,0,360){2}{20}
    \Vertex(10,20){8}
  \end{picture} \\ (a) 
  \end{center}
\end{minipage}
\begin{minipage}[t]{0.47\linewidth}
  \begin{center}
  \begin{picture}(20,60)(0,0)
    \Line(-30,22)(50,22)
    \Line(-30,18)(50,18)
    \Text(-35,22)[r]{$H$}
    \Text(55,22)[l]{$H$}
    \Text(35,60)[br]{$\chi$}
    \ArrowArcn(10,42)(20,270,-90)
    \Vertex(10,20){8}
  \end{picture} \\  (b)
  \end{center}
\end{minipage}
\caption{\it One-loop corrections to the PGB mass in the holographic 
 theory from the gauge field (a) and an elementary fermion (b).
 If the coupling of the source $\chi$ with the conformal sector is
 relevant, then the fermion propagator in the diagram (b) must be 
 intended as corrected by an infinite series of CFT insertions.}
\label{fig:mass}
\end{figure}
At tree level, the Higgs potential vanishes due to the SU(3)$_L$ 
symmetry on the TeV brane. At one-loop level, however, an effective 
potential for $H$ will be induced.  In the holographic picture 
this comes from the interactions between the CFT and the elementary 
fields, which explicitly break the global SU(3)$_L$ symmetry.
At one loop the relevant diagrams for the Higgs mass term are 
those sketched in Fig.~\ref{fig:mass}.  The situation is quite 
similar to QCD, where the charged pion gets a mass at one loop 
due to the fact that the coupling to the photon explicitly breaks 
the global chiral symmetry.  In the 5D picture the effect comes 
from loops of bulk fields that propagate from the TeV brane, where 
$H$ lives, to the Planck brane, where we have the breaking of 
SU(3)$_L$. This is a non-local effect and therefore is finite. 
The one-loop effective potential is similar to that calculated 
in Ref.~\cite{Gherghetta:2003wm}.

The gauge contribution to the Higgs potential arises from loops of 
SU(3)$_L$ gauge bosons. This is given by (the U(1)$_X$ contribution 
can be obtained in a similar way)
\begin{equation}
\begin{split}
  \sqrt{-g_{\rm ind}}\, V_{\rm gauge}(\Sigma) 
    =& \frac{3}{2} \sum_{n=1}^{\infty} \Tr \int_0^\infty 
      \frac{dp}{8\pi^2}\, p^3 \frac{(-1)^{n+1}}{n}
      \Big[{\cal G}\cdot{\cal M}^{2}(\Sigma) \Big]^n 
\\
    =& \frac{3}{2}\Tr \int_0^\infty \frac{dp}{8\pi^2}\, p^3 
      \ln\Big[1+{\cal G}\cdot{\cal M}^2(\Sigma) \Big].
\end{split}
\label{eq:effpotg}
\end{equation}
Here, ${\cal M}^2$ is the boundary squared-mass matrix of the gauge 
boson fields in the background $\Sigma$:
\begin{equation}
  {\cal M}_{ab}^2(\Sigma) 
  = 2g_5^2\, \Sigma^\dagger\, {\rm T}_a{\rm T}_b\, \Sigma,
\label{eq:effmass}
\end{equation}
and ${\cal G}$ is a matrix propagator ${\cal G}_{ab}=G_a(p)\delta_{ab}$,
where $G_a(p)$ are the propagators of the SU(3)$_L$ gauge boson 
($a=1,\cdots,8$) evaluated on the TeV brane with 4D momentum $p$:
$G_a(p)=\hat G(L_1,L_1;p)$ where $\hat{G}(z,z^\prime;p)$ is the 
rescaled gauge boson propagator given in the Appendix with $m_1=0$ 
and $m_0 \sim M_{\rm Pl}$ ($m_0=0$) for the gauge bosons of the 
broken (unbroken) generators on the Planck brane. The effective 
potential of Eq.~(\ref{eq:effpotg}) has the Coleman-Weinberg 
potential form, with the only difference that 4D propagators have 
been replaced by 5D propagators. Plugging Eq.~(\ref{eq:sigmadef}) 
into Eq.~(\ref{eq:effpotg}), we can obtain the gauge contribution 
to the effective potential of $H$. In particular, the mass of $H$ 
for $v L_1 \ll 1$ is given by
\begin{equation}
  m_{H}^2 = \frac{9g_5^2}{4} \left(\frac{L_1}{L_0}\right)^2 
    \int^\infty_0\frac{dp}{8\pi^2}p^3
    \Big[ G_{\rm I\!I\!I}(p)-\frac{2}{3}G_{\rm I\!I}(p)
         -\frac{1}{3}G_{\rm I}(p)\Big],
\label{eq:massH}
\end{equation}
where $G_{\rm I\!I\!I,I\!I,I}(p)$ are the SU(2)$_L$ triplet, doublet 
and singlet components of the SU(3)$_L$ gauge boson propagators. The 
integrand in Eq.~(\ref{eq:massH}) reaches its maximum at $p \sim 1/L_1$
and is exponentially suppressed for $p>1/L_1$. Therefore, $m^2_H$ is 
finite and very insensitive to physics at energies above $1/L_1$.
Notice that, contrary to supersymmetry, the cancellation of quadratic 
divergences in Eq.~(\ref{eq:massH}) is due to particles of equal spin.
Eq.~(\ref{eq:massH}) yields
\begin{equation}
  m_{H}^2 \simeq \left(\frac{0.12}{L_1}\right)^2.
\label{eq:gauge-cont}
\end{equation}
Hence, the PGB mass turns out to be an order of magnitude smaller 
than the scale $1/L_1$, which is also smaller than $\Lambda_{\rm IR}$. 
This mass gap is even larger if $m_H$ is compared with the mass 
of the first KK state, $\Lambda_{\rm NP} \approx \pi/L_1$. 
The contribution of Eq.~(\ref{eq:gauge-cont}) is positive 
and cannot trigger electroweak symmetry breaking by itself.

The top contribution to the effective potential is given by
\begin{equation}
  \sqrt{-g_{\rm ind}}\, V_{\rm top}(\Sigma)
  = -6 \int_0^\infty \frac{dp}{8\pi^2}\, p^3 
    \ln\Big[1+G_i(p)\, m_i^2(\Sigma) \Big],
\label{eq:effpottop}
\end{equation}
where $m_i^2$ is the boundary squared-mass of $Q_L^{(i)}$ in the 
background $\Sigma$:
\begin{equation}
  m_{i}^2(\Sigma) = |\lambda_U|^2\, \Sigma_i^\dagger\Sigma_i,
\label{eq:effmasstop}
\end{equation}
and $G_i(p)$ are the propagators of $Q_L^{(i)}$ ($i=D,S$) 
evaluated on the TeV brane at 4D momentum $p$. These are 
$G_D(p)=\hat{G}_L^{(+,+)}(L_1,L_1;p)$ and 
$G_S(p)=\hat{G}_L^{(-,+)}(L_1,L_1;p)$, where 
$\hat{G}_L^{(\pm,+)}(z,z^\prime;p)$ can be found in the Appendix.
This gives a contribution to the Higgs mass of order 
\begin{equation}
  m^2_H \sim -\frac{h_t^2}{\pi^2} \frac{1}{L_1^2},
\label{eq:top-approx}
\end{equation}
where the exact value depends on $M_Q$.  The top contribution
is negative and, for certain values of $M_Q$, is larger than the 
gauge contribution. The top contribution can then be responsible 
for the breaking of the electroweak symmetry.

Despite the occurrence of electroweak symmetry breaking, the one-loop 
effective potential calculated above cannot by itself lead to a realistic 
scenario of electroweak symmetry breaking. A tree-level Higgs quartic 
coupling is necessary to obtain a large enough physical Higgs mass 
and to generate a Higgs VEV a loop factor smaller than $1/L_1$. In the 
next subsection we will provide a mechanism to generate a Higgs quartic 
coupling, which together with the one-loop gauge and top contributions 
can lead to a realistic theory of electroweak symmetry breaking.

Before concluding this subsection, it is interesting to analyze 
the absence of quadratic divergences in this class of models from 
a 4D perspective. In the standard model, the dominant contribution 
to the Higgs mass term comes from the loop of the top quark and it 
is quadratically divergent. The cancellation of this divergence then 
must arise from a loop of some extra fields. This is indeed what 
happens in our case. The top quark is realized as the zero mode of 
a bulk fermion and it corresponds, in the holographic picture, to 
a mixture between the elementary source and the CFT bound states. 
The physical spectrum then consists of a massless top quark, plus 
a series of CFT bound states that form a complete multiplet of the 
global SU(3)$_L$. It is the contribution of this tower of additional 
states that cancels the divergence of the top loop.  This is, 
therefore, a cancellation involving an infinite numbers of 4D fields.

The picture described above, however, is not quite illuminating to 
understand the finiteness of the top contribution to the Higgs mass. 
To understand it better, we can perform a change of basis going 
from the mass eigenstate basis to the ``interaction basis'', where 
we separate an elementary source field from the tower of composite 
CFT states (the physical top quark is a mixture of these states). 
The whole contribution to the Higgs mass then arises from an exchange 
of the elementary field (Fig.~\ref{fig:mass}), because only the 
elementary sector feels the explicit breaking directly. Now comes 
the most important point. Since the elementary field couples only 
linearly to the CFT sector, the correction to the Higgs mass 
can\textit{not} proceed through a loop of elementary modes 
directly coupled to the Higgs, but it must necessarily involve the 
strong CFT dynamics.  Therefore, a large momentum circulating in 
the top loop always flows into the CFT cloud, and consequently the 
resulting Higgs mass always involves a form factor $F(q^2)$ which 
encodes the non-perturbative effects of the CFT. Since $F(q^2)$ is 
suppressed for $q$ larger than the compositeness scale, the loop 
momentum integral is cut off above that scale.  This is the reason 
why the quadratic divergences are absent in our theory. In this 
picture, no cancellation between different divergent contributions 
is necessary. The Higgs mass correction is finite simply because 
of the form factor suppressing the contribution from large virtual 
momenta: at high energies the constituents of the composite Higgs 
become transparent to the short wavelength probe of the elementary 
fermion, so that $F(q^2)\to 0$ for $q^2\to \infty$. As the explicit 
computation reveals, the damping is exponential. This strong damping 
can be understood by recalling that in the 5D theory the mass 
correction arises as a finite non-local effect. As such, it involves 
a brane to brane propagation, and that explains the exponential 
suppression for internal momenta larger than $\sim 1/L_1$.

A similar scenario, which does not have quadratic divergences for 
the Higgs, was considered in Ref.~\cite{Gherghetta:2003wm}. In that 
model too, the Higgs is a bound state of the CFT, and its tree-level 
mass vanishes since the conformal sector is invariant under a global 
supersymmetry. This invariance is explicitly broken by the interactions 
with the elementary sector, so that the Higgs is expected to acquire 
a mass at one loop. The 5D realization of this scenario consists of 
a warped supersymmetric setup, where matter and gauge fields propagate 
in the bulk of AdS and the Higgs field is localized on the TeV brane. 
If supersymmetry breaking is triggered on the Planck brane, the 
correction to the Higgs mass will be a finite non-local effect.
The corresponding holographic description is almost the same as in 
our case: the Higgs mass is generated through the exchange of some 
elementary field and it is finite because the CFT strong dynamics 
exponentially suppresses contributions from large virtual momenta.

\subsection{A mechanism for the quartic coupling}
\label{subsec:tree-higgs}

Although the Higgs potential is generated by radiative corrections,
it is not sufficient to guarantee a successful phenomenology. We need 
a large $\sim O(1)$ quartic coupling while keeping the quadratic term 
smaller than the effective cutoff scale, {\it i.e.}, the scale that 
suppresses higher dimensional terms in the Higgs potential. Here we 
present an example of dynamics providing such a situation.

The basic idea is simpler to understand in the 4D CFT picture. 
The explicit breaking of SU(3)$_L$ comes from the elementary sector.
Hence, in order to generate a quartic coupling at tree level, the 
Higgs must mix with some elementary scalar $\varphi$. Since the 
mass of the elementary scalar is not protected by any symmetry, 
it is expected to be of order the cutoff scale $\approx k$. 
This implies that, if we want to generate an unsuppressed SU(3)$_L$ 
breaking effect in the Higgs sector, the elementary scalar 
$\varphi$ must be coupled to the CFT with a coupling linear in $k$:
${\cal L}_{\rm int}=\lambda k\, \varphi\cdot {\cal O_\varphi}$, 
{\it i.e.}, the operator ${\cal O_\varphi}$ must have dimension~$2$.
If $\langle 0|{\cal O_\varphi O}_\Sigma^2|0\rangle \neq 0$, where 
${\cal O}_\Sigma$ is an operator that creates $\Sigma$, the Higgs 
will have non-trivial interactions with the scalar $\varphi$, and 
an explicit breaking of SU(3)$_L$ will appear in the Higgs potential 
at tree level. For example, if $\varphi$ is a {\bf 6} of SU(3)$_L$, 
which decomposes under SU(2)$_L$ as a triplet ($\varphi_T$), 
a doublet ($\varphi_D$) and a singlet ($\varphi_S$), a Higgs 
quartic coupling is generated from the diagrams of Fig.~\ref{fig:H-4}. 
We are assuming here that the different SU(2)$_L$ components of 
$\varphi$ have different masses due to the SU(3)$_L$ breaking in 
the elementary sector; otherwise the quartic coupling is zero 
because of the non-linearly realized SU(3)$_L$ symmetry.  In this 
theory, however, there is the danger of also generating a quadratic 
term for $H$. This comes from the diagrams of Fig.~\ref{fig:H-2}. 
A way to avoid the generation of a Higgs squared-mass is to assume 
that the breaking of SU(3)$_L$ in the elementary $\varphi$ affects 
only $\varphi_{T}$ and not the other SU(2)$_L$ components, $\varphi_D$ 
and $\varphi_S$. It is clear that in this case, the diagrams 
of Fig.~\ref{fig:H-2} respect SU(3)$_L$ and vanish. Therefore, 
a mechanism of this type must requires a certain pattern of SU(3)$_L$ 
breaking in the elementary $\varphi$ sector.\footnote{
We could have a Higgs quartic coupling without quadratic term
if only $\varphi_T$ is present. Nevertheless, in this case the 
quartic coupling turns out to be negative.}
\begin{figure}
\begin{minipage}{\linewidth}
\centering
\begin{picture}(0,0)%
\includegraphics{quartic_2.pstex}%
\end{picture}%
\setlength{\unitlength}{1973sp}%
\begingroup\makeatletter\ifx\SetFigFont\undefined%
\gdef\SetFigFont#1#2#3#4#5{%
  \reset@font\fontsize{#1}{#2pt}%
  \fontfamily{#3}\fontseries{#4}\fontshape{#5}%
  \selectfont}%
\fi\endgroup%
\begin{picture}(13275,2251)(901,-1535)
\put(5701,239){\makebox(0,0)[b]{\smash{\SetFigFont{12}{14.4}%
 {\rmdefault}{\mddefault}{\updefault}$H$}}}
\put(5401,-511){\makebox(0,0)[b]{\smash{\SetFigFont{12}{14.4}%
 {\rmdefault}{\mddefault}{\updefault}$H$}}}
\put(5701,-1216){\makebox(0,0)[b]{\smash{\SetFigFont{12}{14.4}%
 {\rmdefault}{\mddefault}{\updefault}$H$}}}
\put(9001,-511){\makebox(0,0)[b]{\smash{\SetFigFont{12}{14.4}%
 {\rmdefault}{\mddefault}{\updefault}$H$}}}
\put(10126,314){\makebox(0,0)[b]{\smash{\SetFigFont{12}{14.4}%
 {\rmdefault}{\mddefault}{\updefault}$H$}}}
\put(10126,-1276){\makebox(0,0)[b]{\smash{\SetFigFont{12}{14.4}%
 {\rmdefault}{\mddefault}{\updefault}$H$}}}
\put(14176,314){\makebox(0,0)[b]{\smash{\SetFigFont{12}{14.4}%
 {\rmdefault}{\mddefault}{\updefault}$H$}}}
\put(14161,-1276){\makebox(0,0)[b]{\smash{\SetFigFont{12}{14.4}%
 {\rmdefault}{\mddefault}{\updefault}$H$}}}
\put(1426,464){\makebox(0,0)[b]{\smash{\SetFigFont{12}{14.4}%
 {\rmdefault}{\mddefault}{\updefault}$H$}}}
\put(901,-136){\makebox(0,0)[b]{\smash{\SetFigFont{12}{14.4}%
 {\rmdefault}{\mddefault}{\updefault}$H$}}}
\put(901,-736){\makebox(0,0)[b]{\smash{\SetFigFont{12}{14.4}%
 {\rmdefault}{\mddefault}{\updefault}$H$}}}
\put(1426,-1441){\makebox(0,0)[b]{\smash{\SetFigFont{12}{14.4}%
 {\rmdefault}{\mddefault}{\updefault}$H$}}}
\put(4201,-451){\makebox(0,0)[b]{\smash{\SetFigFont{10}{12.0}%
 {\rmdefault}{\mddefault}{\updefault}$\times$}}}
\put(3151,-136){\makebox(0,0)[b]{\smash{\SetFigFont{12}{14.4}%
 {\rmdefault}{\mddefault}{\updefault}$\varphi_S$}}}
\put(7651,-136){\makebox(0,0)[b]{\smash{\SetFigFont{12}{14.4}%
 {\rmdefault}{\mddefault}{\updefault}$\varphi_D$}}}
\put(12151,-136){\makebox(0,0)[b]{\smash{\SetFigFont{12}{14.4}%
 {\rmdefault}{\mddefault}{\updefault}$\varphi_{S,T}$}}}
\end{picture}
\caption{\it Contributions to the Higgs quartic coupling. 
 The CFT dynamics is represented as a thick gray line, while 
 a thin black line represents the propagator of the elementary 
 scalar $\varphi$. A cross $\times$ indicates an SU(3)$_L$ 
 breaking by the CFT.}
\label{fig:H-4}
\end{minipage}
\\[0.7cm]
\begin{minipage}{\linewidth}
\centering
\begin{picture}(0,0)%
\includegraphics{quartic_1.pstex}%
\end{picture}%
\setlength{\unitlength}{1973sp}%
\begingroup\makeatletter\ifx\SetFigFont\undefined%
\gdef\SetFigFont#1#2#3#4#5{%
  \reset@font\fontsize{#1}{#2pt}%
  \fontfamily{#3}\fontseries{#4}\fontshape{#5}%
  \selectfont}%
\fi\endgroup%
\begin{picture}(8475,1696)(1201,-1805)
\put(6811,-1111){\makebox(0,0)[b]{\smash{\SetFigFont{12}{14.4}%
 {\rmdefault}{\mddefault}{\updefault}$H$}}}
\put(1201,-361){\makebox(0,0)[b]{\smash{\SetFigFont{12}{14.4}%
 {\rmdefault}{\mddefault}{\updefault}$H$}}}
\put(1201,-1711){\makebox(0,0)[b]{\smash{\SetFigFont{12}{14.4}%
 {\rmdefault}{\mddefault}{\updefault}$H$}}}
\put(9676,-1111){\makebox(0,0)[b]{\smash{\SetFigFont{12}{14.4}%
 {\rmdefault}{\mddefault}{\updefault}$H$}}}
\put(4201,-1036){\makebox(0,0)[b]{\smash{\SetFigFont{10}{12.0}%
 {\rmdefault}{\mddefault}{\updefault}$\times$}}}
\put(3151,-736){\makebox(0,0)[b]{\smash{\SetFigFont{12}{14.4}%
 {\rmdefault}{\mddefault}{\updefault}$\varphi_S$}}}
\put(8251,-736){\makebox(0,0)[b]{\smash{\SetFigFont{12}{14.4}%
 {\rmdefault}{\mddefault}{\updefault}$\varphi_D$}}}
\end{picture}
\caption{\it Contributions to the Higgs mass. The CFT dynamics is 
 represented as a thick gray line, while a thin black line represents 
 the propagator of the elementary scalar $\varphi$. A cross $\times$ 
 indicates an SU(3)$_L$ breaking by the CFT.}
\label{fig:H-2}
\end{minipage}
\end{figure}

Let us see how this idea is implemented in the 5D AdS picture. 
By AdS/CFT, a scalar operator of dimension $2$ corresponds to a bulk 
scalar $\Phi$ of squared-mass $M^2_{\Phi}=-4k^2$. Note that as long as 
$M_\Phi \geq -4k^2$ (and certain conditions for the brane masses are 
met), $\Phi$ is not tachyonic and does not develop a VEV. The 5D scalar 
$\Phi$ is responsible for communicating the SU(3)$_L$ breaking {\it on 
the Planck brane} to the TeV brane where the Higgs lives.\footnote{
Another possibility, which could avoid introducing $\Phi$, is 
to have a Higgs with a profile in the extra dimension.}
The field $\Phi$ must be coupled to the Higgs on the TeV brane.
We thus require $\Phi$ to transform as ${\bf 6_{2/3}}$ under 
SU(3)$_L \times$U(1)$_X$ and have the following coupling:
\begin{equation}
  {\cal L} = \delta(z-L_1) \sqrt{-g_{\rm ind}} 
  \left[ \lambda_\Phi \Sigma \Phi^\dagger \Sigma + {\rm h.c.} \right],
\end{equation}
where $\lambda_\Phi$ is a coupling of mass dimensions $1/2$. A small 
deviation from $M^2_\Phi = -4k^2$, as arising for example from the 
one-loop correction to the bulk mass of $\Phi$, will not spoil 
our mechanism.

By integrating out the $\Phi$ field, we find a tree-level 
Higgs potential in the low-energy effective theory,
\begin{equation}
  V_{H,{\rm tree}} = m_H^2 |H|^2 + \frac{\lambda_H}{2} |H|^4.
\label{eq:efff}
\end{equation}
$m^2_H$ and $\lambda_H$ are generated by 5D diagrams similar
to those in Fig.~\ref{fig:H-2} and Fig.~\ref{fig:H-4} respectively, 
where the internal propagators now represent the 5D field $\Phi$ 
with the appropriate SU(2)$_L$ quantum numbers. We find
\begin{gather}
  m_H^2 = 2 \lambda_\Phi^2 v^2 \left[\hat{G}_S-\hat{G}_D\right],
\label{eq:H2-1} \\
  \lambda_H = \frac{2}{3} \lambda_\Phi^2 
    \left[ 8\hat{G}_D - 5\hat{G}_S - 3\hat{G}_T \right],
\label{eq:H4-1}
\end{gather}
where $\hat{G}_S$, $\hat{G}_D$ and $\hat{G}_T$ are the rescaled 
propagators of the SU(2)$_L$ singlet, doublet and triplet components 
of $\Phi$ with end points on the TeV brane, evaluated at zero 4D momentum.
The explicit form of these propagators can be found in the Appendix. 
For $M_\Phi^2 \simeq - 4k^2$, which we assume here, they are given by
\begin{equation}
  \hat{G}_a \simeq \frac{1}{r_a}
    \left[ 1 + (m_{0,a}L_0-2)\ln\left(\frac{L_1}{L_0}\right) \right],
\label{eq:prop-TeV}
\end{equation}
where $r_a \equiv m_{0,a}+m_1+L_0(m_{0,a}-2L_0^{-1})(m_1+2L_0^{-1})
\ln(L_1/L_0)$ and $a = S,D,T$. We introduced a common scalar mass 
$m_1$ on the TeV brane for the SU(2)$_L$ singlet, doublet and triplet 
components of $\Phi$ field, but allowed different masses $m_{0,a}$ on 
the Planck brane. The values of the masses $m_{0,a}$ are determined by 
the high-energy SU(3)$_L$-breaking dynamics on the Planck brane. 
Here we do not specify a particular pattern for this symmetry breaking, 
but rather we look for the parameter region of $m_{0,a}$ (and $m_1$) 
where the successful Higgs phenomenology is obtained.

Inserting Eq.~(\ref{eq:prop-TeV}) into Eq.~(\ref{eq:H2-1}), we obtain 
\begin{equation}
  m_H^2 \simeq \frac{2 \lambda_\Phi^2}{r_D\, r_S} (m_{0,D}-m_{0,S})\, v^2.
\label{eq:H2-2}
\end{equation}
We see that, as expected, if the doublet and singlet components of 
$\Phi$ do not feel the breaking of SU(3)$_L$ on the Planck brane, 
$m_{0,D}= m_{0,S}$, the resulting Higgs squared-mass parameter is zero. 
Assuming this, the quartic coupling, Eq.~(\ref{eq:H4-1}), is given by
\begin{equation}
  \lambda_H \simeq \frac{2 \lambda_\Phi^2}{r_T\, r_S} (m_{0,T}-m_{0,S}).
\label{eq:H4-2}
\end{equation}
Therefore, for $m_{0,T} \simgt m_{0,S}$, we obtain a sufficiently 
large Higgs quartic coupling.

We have seen that the desirable Higgs potential is obtained for 
$m_{0,T} \simgt m_{0,S} \simeq m_{0,D}$.\footnote{
An alternative parameter region is $m_{0,D},m_{0,S} \simlt k \simlt 
m_{0,T},m_1$, in which case the tree-level Higgs potential is given 
by $m_H^2 \simeq \lambda_\Phi^2 (m_{0,D}-m_{0,S}) (2 \ln(L_1/L_0)^2 
m_1^2)^{-1} v^2$ and $\lambda_H \simeq 2 \lambda_\Phi^2 m_{0,T} / 
(m_1 r_T)$ so that we can have $m_H^2 \ll v^2$, $\lambda_H \sim 1$.}
Although this may appear a rather ad hoc hypothesis, we stress that 
it is an assumption about the underlying physics at the Planck scale 
which is responsible for the symmetry breaking. We do not explicitly 
address this sector here, but we expect that there are some mechanisms 
realizing (approximately) the situation discussed above. We must say, 
however, that even if $m_{0,S} = m_{0,D}$ at tree level due to some 
specific SU(3)$_L$ breaking pattern on the Planck brane, quantum 
effects (coming, for example, from gauge interactions) will modify 
this relation. Therefore, a Higgs squared-mass will be induced from 
Eq.~(\ref{eq:H2-2}), at least, at the quantum level. This contribution 
is difficult to calculate since it depends on the Planck-brane 
fields that break SU(3)$_L$, but it can be estimated to be one-loop 
factor smaller than $v^2$. We can then conclude that the Higgs 
potential Eq.~(\ref{eq:efff}) together with the one-loop contributions 
of Eqs.~(\ref{eq:effpotg}) and (\ref{eq:effpottop}) can lead to 
a successful electroweak symmetry breaking with a VEV for the Higgs 
a loop factor smaller than $1/L_1$, and a physical Higgs mass 
larger than the experimental bound. The precise determination 
of $\langle H \rangle$ and $m_h$ is, however, not possible here 
due to the dependence of $\lambda_H$ on the unknown free parameters 
of the model.

\section{Pseudo-Goldstone Bosons as Holograms of $A_5$}
\label{sec:A5higgs}

In this section we consider the possibility of breaking the gauge
symmetry in the warped extra dimension by boundary conditions, 
and not by scalar fields on the two branes. We will show that the 
holographic picture is almost the same as before and that the 
holographic PGB in this case corresponds in the 5D theory to the 
zero mode of the fifth component of the gauge boson, $A_5$. We 
note that the model presented in the previous section and the one 
presented here must coincide in the limit $v \gg \Lambda_{\rm IR}$, 
since the breaking of SU(3)$_L$ by the scalar $\Sigma$ is 
equivalent to a breaking by boundary conditions in the limit 
$v \rightarrow \infty$~\cite{Nomura:2001mf}.

Let us consider the most general case of a bulk gauge group $G$ 
reduced to the subgroups $H_0$ and $H_1$ on the Planck and TeV branes, 
respectively. This corresponds to assigning the following boundary 
conditions to the gauge bosons at the Planck and TeV branes:
\begin{equation}
\begin{split}
  A_\mu^a       \;&(+,+)\qquad\quad T^a        \in\mathsf{Alg}\{H\},     \\
  A_\mu^{\bar a}\;&(+,-)\qquad\quad T^{\bar a} \in\mathsf{Alg}\{H_0/H\}, \\
  A_\mu^{\dot a}\;&(-,+)\qquad\quad T^{\dot a} \in\mathsf{Alg}\{H_1/H\}, \\
  A_\mu^{\hat a}\;&(-,-).
\end{split}
\label{eq:parities}
\end{equation}
Here, by $+$ ($-$) we denote the Neumann (Dirichlet) boundary condition, 
and $H=H_0 \cap H_1$. The $A_5$'s have the opposite boundary conditions 
to those of the corresponding $A_\mu$'s.

The holographic 4D theory consists of a CFT sector whose global 
invariance $G$ is spontaneously broken down to $H_1$ by strong 
dynamics, with an order parameter of $\mathcal{O}({\rm TeV})$.
External gauge fields weakly gauge the subgroup $H_0$ of $G$:
\begin{equation}
  {\cal L} = {\cal L}_\text{CFT} - \frac{1}{4g^2} (F_{\mu\nu}^\alpha)^2 
    + A_\mu^\alpha J^{\mu \,\alpha}, \quad\quad \alpha=a,\bar{a}.
\label{eq:gauging}
\end{equation}
This situation is somewhat different from the model of 
section~\ref{sec:higgsonb}, where the whole $G$ was gauged in 
the 4D theory and Higgsed down to $H_0$ at high energies. The two 
scenarios, however, are indistinguishable from low-energy observers.
The gauging of only a subgroup of the global symmetry $G$ is 
experienced by the CFT as an explicit breaking of $G$.

\begin{figure}[t]
\begin{minipage}{0.54\linewidth}
  \centering
  \begin{picture}(200,140)
  \SetWidth{1} \GCirc(60,60){48}{0.85} \SetWidth{0.5}
  \Vertex(95,75){2} \Line(95,75)(135,90) \Text(142,93)[l]{$\varphi$}
  \Vertex(85,90){2} \Photon(85,90)(135,120){2.5}{4.5} 
  \Text(142,123)[l]{$A_\mu^a\, ,A_\mu^{\bar a}$}
  \Text(60,75)[c]{\textsc{\Large CFT}}
  \Text(60,45)[c]{\large $G\to H_1$} 
  \Text(60,27)[c]{$\pi^{\hat a},\pi^{\bar a}$}
  \end{picture}
\caption{\it The holographic theory consists of a CFT interacting with 
 an elementary sector represented here by the gauge fields $A_\mu^a$, 
 $A^{\bar a}_\mu$ and a generic field $\varphi$. The Goldstone bosons 
 $\pi^{\bar a}$ are eaten by the gauge fields $A_\mu^{\bar a}$ to form 
 massive vectors; the remaining $\pi^{\hat a}$ are PGBs.}
\label{fig:holo}
\end{minipage}
\hspace{0.5cm}
\begin{minipage}{0.40\linewidth}
  \centering
  \begin{picture}(140,140)
  \CArc(70,70)(60,0,360) \CArc(55,60)(33,0,360) \CArc(85,60)(33,0,360) 
  \Text(70,70)[c]{$H$} \Text(70,55)[c]{\scriptsize $(+,+)$} 
  \Text(70,110)[c]{\scriptsize $(-,-)$} \Text(125,110)[l]{$G$}
  \Text(37,70)[c]{$H_0$} \Text(37,55)[c]{\scriptsize $(+,-)$}
  \Text(103,70)[c]{$H_1$} \Text(103,55)[c]{\scriptsize $(-,+)$}
  \end{picture}
\caption{\it The pattern of symmetry breaking.}
\label{fig:breaking}
\end{minipage}
\end{figure}
Let us count the number of PGBs present in the theory. The spontaneous 
breaking in the CFT sector delivers $n=\text{dim}(G/H_1)$ Goldstone 
bosons.  However, the gauging of $H_0$, Eq.~(\ref{eq:gauging}), 
makes part of them, $m=\text{dim}(H_0/H)$, being eaten by the gauge 
bosons $A_\mu^{\bar a}$. The remaining $n-m$ are PGBs; they are 
massless at tree level, but they acquire masses from radiative 
corrections due to the explicit breaking of $G$ by the interaction 
terms of Eq.~(\ref{eq:gauging}).  Only the gauge bosons associated 
to the symmetry subgroup $H=H_0 \cap H_1$ are exactly massless. 
Figs.~\ref{fig:holo} and \ref{fig:breaking} give a pictorial 
representation of the holographic theory and of the symmetry 
breaking pattern.

In general, any interaction with the elementary sector can 
communicate the explicit breaking of $G$ to the CFT sector. 
The AdS/CFT correspondence prescribes that adding a generic field 
in the bulk of AdS with Neumann boundary conditions on the Planck 
brane corresponds to modifying the CFT content and adding some 
elementary 4D field $\varphi$ which couples to the conformal 
sector through a coupling $\varphi \cdot {\cal O}_\varphi$ (see 
Fig.~\ref{fig:holo}). Since $\varphi$ will come in a representation 
of the group $H_0$ (rather than $G$), which is the symmetry of the 
elementary sector, the coupling $\varphi \cdot {\cal O}_\varphi$ 
is not $G$-invariant.  As the only source of explicit breaking is 
represented by those interactions, this necessarily implies that 
the mass terms for the PGBs are generated only through processes 
where elementary fields are exchanged (Fig.~\ref{fig:mass}).

Since the 4D holographic picture and the 5D AdS setup describe the 
same physics, they must exhibit the same physical spectrum. Therefore, 
there must be $n-m$ massless scalars in the 5D theory after KK reduction. 
We now see in detail how these massless scalars appear. The 5D gauge 
Lagrangian is given by
\begin{equation}
  {\cal L}_{\rm gauge} 
  = \sqrt{g}\, \Big[ -\frac{1}{4 g_5^2}\, g^{KM}g^{LN} F_{KL}F_{MN}
  + {\cal L}_{\rm GF} \Big],
\label{eq:5dlag}
\end{equation}
where the metric $g_{MN}$ is that of Eq.~(\ref{eq:metric}) and 
${\cal L}_{\rm GF}$ is the gauge-fixing term. A convenient choice 
for ${\cal L}_{\rm GF}$ is
\begin{equation}
  {\cal L}_{\rm GF} 
  = -\frac{1}{2\xi g_5^2} \Big[g^{\mu\nu} \partial_\mu A_\nu 
    + z\, \xi\, g^{55}\, \partial_z (A_5/z) \Big]^2,
\end{equation}
with which all mixing terms between $A_\mu$ and $A_5$ 
cancel~\cite{Randall:2001gb}.  Eq.~(\ref{eq:5dlag}) 
can be written as 
\begin{equation}
\begin{split}
  {\cal L}_{\rm gauge} 
  = \frac{1}{2 g_5^2 kz}\, 
    \bigg\{& A_\mu \big[ \eta^{\mu\nu}\partial_\rho\partial^\rho 
    - (1-1/\xi)\,\partial^\mu\partial^\nu \big]A_\nu 
    + (\partial_z A_\nu)^2 
\\
  & + (\partial_\mu A_5)^2- \xi \, z^2 
    \left(\partial_z \frac{A_5}{z}\right)^2 \bigg\} + \cdots,
\end{split}
\label{eq:RSLg2}
\end{equation}
where the ellipses stand for interaction terms; 4D indices are 
raised/lowered with $\eta_{\mu\nu}$. We now perform a KK reduction.
We are interested only in the massless scalar spectrum.  This comes 
from the zero modes of the fifth components of the gauge bosons, 
$A_5(x,z)=f_0(z)A_5^{(0)}(x)+\cdots$, where $f_0(z)$ satisfies 
\begin{equation}
  \partial_z \left(\frac{f_0(z)}{z}\right) = 0.
\end{equation}
The solution to this equation is only compatible with $(+,+)$ boundary 
conditions, in which case we obtain
\begin{equation}
  f_0(z) = \frac{z}{N_0}, \ \ \ \ \ 
    {\rm where}\ \  N_0=\sqrt{\frac{L_1^2-L_0^2}{2}}.
\label{eq:wavefun}
\end{equation}
The components of $A_5$ with $(+,+)$ boundary conditions are $A_5^{\hat a}$.
There are ${\rm dim}[G/H_1]-{\rm dim}[H_0/H]$ of them, as expected from 
the 4D dual picture.  Therefore the massless modes of $A_5^{\hat a}$ must 
correspond to the PGBs of the 4D CFT~\cite{Contino:2002kc}. A further 
check of this correspondence comes from their wavefunction. It is peaked 
towards the TeV brane as expected if the holographic PGBs are really 
bound states of the CFT. The excited KK modes of $A_5$ can be eliminated 
from the spectrum by going to the unitary gauge $\xi=\infty$.

It is clear from the 5D point of view that a tree-level potential 
for $A_5$ is absent, because it is forbidden by gauge invariance. 
An effective potential is then generated radiatively as 
a function of the non-local, gauge invariant Wilson line 
$W=\text{Tr}\,\mathcal{P}\exp\big(i\int_{L_1}^{L_0}\! dz\, A_5\big)$. 
This implies that $A_5$ will get a mass at one loop, which, by 
non-locality, is finite and cutoff insensitive.  In general, the 
energy is minimized at a nonzero background value of $A_5$, triggering 
spontaneous breaking of the symmetry $H$~\cite{Hosotani:1983xw}.
As in flat space, different values of the background ${\cal A}_5 = f_0(z) 
c^{\hat a}T^{\hat a}$, with $c^{\hat a}={\rm const.}$, define physically 
inequivalent vacua. This is because one cannot find a continuous 
gauge parameter $\theta(z)=\theta^{\hat a}(z)T^{\hat a}$ with the 
defined boundary conditions $(-,-)$ that eliminates ${\cal A}_5$ 
by the use of the gauge transformation ${\cal A}_5 \rightarrow 
U {\cal A}_5 U^\dagger + U \partial_z \theta\, U^\dagger$. 
Nevertheless, by relaxing one of the two boundary conditions, 
{\it e.g.} that on the Planck brane, it is possible to ``gauge away'' 
${\cal A}_5$ by the transformation with
\begin{equation}
 \theta(z) = -\int^z_{L_1}\!dz\, {\cal A}_5.
\end{equation}
Under this gauge transformation, however, the charged bulk fields are 
also transformed: $\Phi(x,z) \rightarrow e^{i\theta(z)} \Phi(x,z)$.
This implies that the theory with ${\cal A}_5=0$ is equivalent to that 
with nonzero background ${\cal A}_5$, but only if we use the redefined 
bulk fields $e^{i\theta(z)}\Phi \equiv \Phi'$ instead of $\Phi$.
The Planck-brane boundary conditions of the fields $\Phi'$ are then 
different from those of the fields $\Phi$, since $\Phi$ and $\Phi'$ 
differ at $z=L_0$ by a non-trivial gauge phase:
\begin{equation}
  \Phi'(L_0) = e^{-i\theta(L_0)} \Phi(L_0).
\label{eq:cbc}
\end{equation}
This phase is the Wilson line.

The one-loop contributions to the effective potential of $A_5$ are 
easily estimated as follows. The appearance of the Wilson line 
in the vacuum energy requires a contribution of a bulk field that 
propagates from one brane (at $L_0$) to the other (at $L_1$). 
The energy involved in this contribution is then of the order 
of the inverse of the conformal distance between the branes, 
$E \sim 1/\int^{L_1}_{L_0}dz \sim 1/L_1$.  Therefore, the 
mass of $A_5$ is estimated to be $m^2_{A_5} \sim g^2_5 (L_1/L_0)^2 
E^4\, \hat{G}(L_0,L_1;p)|_{E\sim p\sim 1/L_1}$, where $\hat{G}(z,z';p)$ 
is the propagator of the bulk field given in the Appendix.  It is 
interesting to look at the limit $L_0 \rightarrow 0$. In this case 
the propagators from $L_0$ to $L_1$ for the gauge boson and the 
graviton vanish, implying that no effective potential is induced 
for $A_5$. We thus find that $A_5$'s are massless in this limit 
at all loop orders.  Notice that, contrary to the flat space case, 
the zero modes of $A_5$ are still normalizable modes even though the 
extra dimension is infinite [see Eq.~(\ref{eq:wavefun})], and thus 
remain in the theory as massless scalars.  This is in fact what we 
expect from holography. In the 4D picture, the limit $L_0\rightarrow 0$ 
corresponds to sending the UV cutoff to infinity. This implies that 
the 4D low-energy gauge coupling becomes zero (gaugeless limit), and 
the gauge bosons that explicitly break $G$ decouple from the theory, 
making the PGBs true Goldstone bosons. Note that in the gaugeless limit 
the number of Goldstone bosons is $n=\text{dim}(G/H_1)$ instead of $n-m$. 
In the 5D AdS the $m$ extra massless scalars come from $A_5^{\bar a}$;
they have $(-,+)$ boundary conditions and admit zero modes for $L_0=0$.

The limit $L_0 \to 0$ is subtler when other interactions are 
present. The point is that interactions between the CFT and the 
elementary sector that proceed through a {\it relevant} coupling 
are not expected to die off when the UV cutoff goes to infinity.
This is the case, for example, of the interaction between 
an elementary chiral fermion $\chi$ and a CFT operator, 
${\cal L}_{\rm int}=\bar{\chi} {\cal O}$, for ${\rm dim}[{\cal O}]<5/2$.
By AdS/CFT this corresponds to a bulk fermion with mass $|M_\Psi|<k/2$.
In the 5D picture the non-decoupling is evident from the fact that 
the brane to brane fermionic propagator does not go to zero in the 
limit $L_0 \rightarrow 0$ for $|M_\Psi|<k/2$.

\subsection{The mass of $A_5$ at one loop}
\label{subsec:mass-comp}

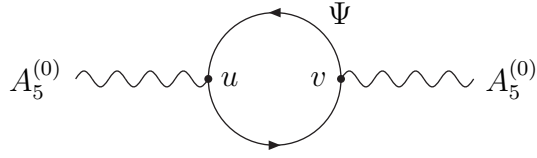
\begin{figure}[t]
\begin{center}
   \begin{picture}(150,50)
   \ArrowArc(75,25)(25,0,180) \ArrowArc(75,25)(25,180,360)
   \Photon(0,25)(50,25){3}{4} \Photon(100,25)(150,25){3}{4}
   \Vertex(50,25){1.5} \Vertex(100,25){1.5} 
   \Text(55,25)[l]{$u$} \Text(95,25)[r]{$v$} \Text(100,50)[c]{$\Psi$}
   \Text(-5,25)[r]{$A_5^{(0)}$} \Text(155,25)[l]{$A_5^{(0)}$}
   \end{picture}
\end{center}
\caption{\it One-loop correction in AdS to the $A_5$ zero-mode mass term.}
\label{fig:5dloop}
\end{figure}

We present here the calculation of the mass of $A_5^{(0)}$ at one-loop 
level, which confirms the statements made above. We consider the simple 
case of Eq.~(\ref{eq:parities}) with $H_0=H_1=H$, and concentrate 
on the contribution from a bulk fermion with a 5D mass $M_\Psi$ 
and the following boundary conditions:
\begin{equation}
\begin{split}
  A_\mu^a       \; &(+,+), \quad A_5^a \;       (-,-),\\
  A_\mu^{\hat a}\; &(-,-), \quad A_5^{\hat a}\; (+,+),
\end{split}
\hspace{1.5cm}
  \Psi = \begin{bmatrix} \psi^i_L (+,+) & \psi^i_R (-,-) \\ 
  \psi^{\textit{\^\i}}_L (-,-) & \psi^{\textit{\^\i}}_R (+,+) 
  \end{bmatrix}.
\label{eq:parit}
\end{equation}
The relevant diagram at one loop is depicted in Fig.~\ref{fig:5dloop}.
The contribution from other particles can be easily derived from 
this result.

The Lagrangian of a 5D fermion with a constant bulk mass $M_\Psi$ is
\begin{equation}
\label{eq:Lfermion}
  {\cal L} = \sqrt{g}\, 
    \left[ \frac{i}{2}\, \bar{\Psi} e^M_A \Gamma^A D_M \Psi
    - \frac{i}{2}\, (D_M \Psi)^\dagger \Gamma^0 e^M_A \Gamma^A D_M\Psi 
    - M_\Psi\bar{\Psi}\Psi \right],
\end{equation}
with $e^M_A=kz\,\delta^M_A$ the inverse vielbein and 
$\Gamma^M=\{ \gamma^\mu,-i\gamma^5 \}$ the 5D Dirac matrices.
The covariant derivative is
\begin{equation}
  D_M = \partial_M + \frac{1}{8} \omega_{M\, AB} 
    \left[ \Gamma^A, \Gamma^B \right] - i A_M,
\end{equation}
where the only non-vanishing entries in the spin connection 
$\omega_M^{AB}$ are $\omega_{\mu\, a5}=-\eta_{\mu a}/z$.
The mass correction to the zero mode of $A_5$ is written as
\begin{equation}
\begin{split}
  m_{A_5}^2 = - (g_5^2 k)\, C(r) 
    \int \frac{d^4p}{(2\pi)^4} \int_{L_0}^{L_1} 
    & du\, \frac{1}{(ku)^4} \int_{L_0}^{L_1} dv\, 
    \frac{1}{(kv)^4}\, f_0(u)f_0(v) \\
  &\times \Tr\left[ \gamma^5\, iS^{(+,+)}(v,u;p)\,
    \gamma^5\, iS^{(-,-)}(u,v;p)\right],
\end{split}
\label{eq:intexp}
\end{equation}
where $C(r)$ is the Dynkin index $\Tr(T^\alpha T^\beta)=C(r)\, 
\delta^{\alpha\beta}$ for a fermion in the representation $r$.
Here, $f_0$ is the $A_5$ zero-mode wavefunction, Eq.~(\ref{eq:wavefun}), 
and $S^{(\pm,\pm)}(z,z^\prime;p)$ denotes the propagator of a 5D 
fermion, with boundary conditions $(\pm,\pm)$ and 4D momentum $p$, 
between the two points $z$ and $z^\prime$ along the fifth dimension. 

Using the fermion propagator given in the Appendix, one can obtain 
an expression for $m_{A_5}^2$ in terms of integrals of Bessel 
functions. In the particular case of integer values of $M_\Psi/k$, 
the Bessel functions reduce to trigonometric functions so that 
the integrals greatly simplify.  For example, when $M_\Psi=0$ 
Eq.~(\ref{eq:intexp}) becomes, after some algebra:
\begin{equation}
  m_{A_5}^2 = -\frac{C(r)}{\pi^2} (g_5^2 k) 
    \int_{L_0}^{L_1}\! dv\; f_0(v) \int_{L_0}^{v}\! du\; f_0(u) 
    \int_0^\infty\! dp\; \frac{p^3}{\sinh [p(L_1-L_0)]}.
\end{equation}
The integrals are convergent and the result is finite. The momentum 
integral involves the brane to brane propagator, as expected 
from the previous discussion, and it converges exponentially. 
Performing the integrals one obtains the result
\begin{equation}
  m^2_{A_5} = - \frac{C(r)}{\pi^2}\, (g_5^2 k)\, 
    \frac{1}{L_1^2-L_0^2} \, F(L_0/L_1),
\label{eq:result}
\end{equation}
where the function $F(x)$ is given, for example, by
\begin{equation}
  F(x)|_{M_\Psi=0} = \frac{3}{8}\, \zeta(3)\, \frac{(1+x)^2}{(1-x)^2},
\end{equation}
\begin{equation}
\begin{split}
  F(x)|_{M_\Psi=k} &= \frac{x (1+x)^2}{4(1-x)^2} \; 
    \int_0^\infty\! dt \; \frac{t^5}{\sinh t}\, 
    \frac{1}{\big[(x-1)^2\, t\,\cosh t 
    + \left(-1+x(2-x+t^2)\right)\sinh t \big]}
\\
  &\simeq 1.67\, x + \mathcal{O}(x^2),
\end{split}
\end{equation}
\begin{equation}
\begin{split}
  F(x)|_{M_\Psi=2k} =& \frac{x^3 (1+x)^2}{4(1-x)^2}\; 
    \int_0^\infty\! dt\; \frac{t^9}{\big[(x-1)^2\, t\,\cosh t
    + \left(-1+x(2-x+t^2)\right)\sinh t \big]}
\\
  & \times \Big[3 (1-x)^2 \, (t^3+6t-3xt-3)\, t\, \cosh t
\\
  &\qquad + \big(9(1-x)^4
    + 3(1-x)^2 (1+x^2-3x)t^2+x^2 t^4\big)\sinh t\Big]^{-1}
\\
 \simeq &12.4\, x^3 + \mathcal{O}(x^4),
\end{split}
\end{equation}
for the case of $M_\Psi/k=0,1,2$, respectively. We find that the $A_5$ 
mass correction is $\mathcal{O}(1/L_1^2)$ for $|M_\Psi|<k/2$, while 
it receives a strong suppression for $|M_\Psi|>k/2$ ($F(x)|_{M_\Psi=ck} 
\propto x^{2|c|-1}$ for $|c| > 1/2$).  This is in agreement with the 
holographic picture where for $|M_\Psi|>k/2$ the CFT operator coupled 
to the elementary fermion becomes irrelevant and the Yukawa coupling 
becomes small as shown in Eq.~(\ref{eq:ht}).\footnote{
Apparently, Eq.~(\ref{eq:result}) does not seem to give a one-loop 
suppression of $m_{A_5}^2$ compared with $1/L_1^2$ for $M_\Psi < k/2$, 
as $g_5^2 k$ is expected to be large $\sim \ln(L_1/L_0)$ in realistic 
cases. This is because the 4D Yukawa coupling is large $\sim 
\sqrt{g_5^2 k}$ for $M_\Psi < k/2$, canceling the suppression from 
the loop factor.  On the other hand, for $M_\Psi = k/2$ the 4D Yukawa 
coupling is given by $\sqrt{g_5^2 k/\ln(L_1/L_0)} = O(1)$, so that 
$m_{A_5}^2$ is one-loop suppressed compared with $1/L_1^2$. In general, 
written in terms of the 4D Yukawa coupling $h_t$, Eq.~(\ref{eq:result}) 
always yields the result Eq.~(\ref{eq:top-approx}) regardless of 
the value of $M_\Psi$, i.e., the value of $h_t$.}

It is interesting to notice that $m_{A_5}^2$ is even under a change 
of the sign of $M_\Psi$. From a 5D perspective this is expected, 
as a change of the sign in the bulk fermion mass is equivalent to 
inverting the chirality, $L \leftrightarrow R$. Given the assignment 
for the boundary conditions of the fermion, Eq.~(\ref{eq:parit}), 
this chirality inversion corresponds to exchanging the $i$ superscript 
with $\textit{\^\i}$, an operation which leaves Eq.~(\ref{eq:result}) 
invariant. From a 4D holographic perspective, on the other hand, the 
fact that the result does not depend on the sign of $M_\Psi$ arises 
as a consequence of the requirement that the two CFT descriptions in 
terms of the left-handed and right-handed sources, Eq.~(\ref{eq:holo1}) 
and Eq.~(\ref{eq:holo2}), are equivalent for $-k/2 \leq M_\Psi \leq k/2$.

\subsection{The standard-model Higgs as a hologram of $A_5$}

It has already been noticed~\cite{Arkani-Hamed:2001nc} that the 
Higgs as a PGB can be realized as the discretization of the Wilson-loop 
in a deconstructed fifth dimension.  We have shown here that the 
connection can be even more strict: the PGB can be the holographic 
image of the fifth component of the gauge field that lives in 
a warped extra dimension. In fact, such a Higgs happens to be 
a composite bound state of a strongly interacting (conformal) 
sector, so that different ideas, which previously seemed distinct, 
merge together into a single scenario. Moreover, if the bulk group 
$G$ is simple, one could also pursue the idea of unification of the 
standard-model electroweak interactions.  If boundary gauge kinetic 
terms do not play a major role, the renormalization-group flow of 
the gauge couplings, in the general scenario of Eq.~(\ref{eq:parities}), 
will follow the pattern:
\begin{center}
\begin{picture}(400,40)(0,0)
  \Text(40,35)[c]{$H=H_0\cap H_1$} \LongArrow(80,35)(180,35) 
  \Text(200,35)[c]{$H_1$} \LongArrow(220,35)(320,35) \Text(340,35)[c]{$G$}
  \Text(40,8)[c]{\parbox[c]{40 pt}{\centering at low \\ energy}} 
  \Text(200,8)[c]{\parbox[c]{90 pt}{\centering exact unification 
    \\ at $E\sim 1/L_1$}}
  \Text(340,8)[c]{\parbox[c]{130 pt}{\centering apparent ``unification'' 
    \\ at $E\sim 1/L_0$}}
\end{picture}
\end{center}
Strictly speaking, the gauge couplings of the holographic theory 
never become $G$-symmetric at high energies.  Nevertheless, low-energy 
observers still find unification predictions from $G$ as if unification 
occurred at $E \sim 1/L_0$~\cite{Goldberger:2002pc}, because the gauge 
couplings in the holographic theory become strong at $E \sim 1/L_0$ 
and thus their low-energy values are insensitive to the initial values 
at high energies.

The most economical choice for a simple electroweak group is SU(3)$_L$.
In this case it is known that the hypercharge normalization does 
not come out correct: embedding the Higgs in an adjoint of SU(3)$_L$ 
gives a prediction $\sin^2\theta_W=3/4$~\cite{Manton:1979kb}, which 
cannot be accommodated neither with a unification at TeV nor 
at $M_{\rm Pl}$. This difficulty, however, is avoided if we introduce 
brane-localized gauge kinetic terms such that the low-energy gauge 
coupling values are correctly reproduced. In this case the above 
picture is modified by large threshold corrections arising either 
at $E \sim 1/L_0$ or $1/L_1$, depending on where we put the 
brane-localized kinetic terms.  The situation is similar in the 
case where one tries to embed also the standard-model SU(3)$_C$  
into the bulk simple group. The simplest possibility in this case 
is SU(6). Although a naive prediction from the SU(6) group theory 
does not yield the observed values of the gauge couplings, we can 
always adjust them by appropriately choosing the coefficients of 
brane-localized gauge kinetic terms, which do not necessarily 
respect SU(6).  Therefore, although these theories are not as 
predictive as 4D supersymmetric unified theories, they are not 
in contradiction with the observed value of the low-energy gauge 
couplings.

An important issue for any realistic theory with the Higgs as $A_5$ 
is to have correct Yukawa couplings between matter and the Higgs. This 
issue was studied in flat space in Ref.~\cite{Burdman:2002se}, and the 
mechanism considered there can be applied in our warped case without 
any essential modification. For example, we can adopt the SU(6) model 
of~\cite{Burdman:2002se} (either supersymmetric or non-supersymmetric) 
with the gauge group reduced to SU(5)$\times$U(1)$_X$ on the Planck 
brane and to SU(4)$_C \times$SU(2)$_L \times$U(1) on the TeV brane.
Below we explain things in the supersymmetric case. The matter fields 
are introduced as hypermultiplets, transforming as ${\cal D}({\bf 15})$, 
${\cal U}({\bf 20})$ and ${\cal E}({\bf 15})$ (and ${\cal N}({\bf 6})$ 
for right-handed neutrinos) under SU(6), together with some brane fields. 
[In the non-supersymmetric case, these fields are bulk and brane fermions.] 
Realistic Yukawa matrices are then reproduced by appropriately choosing 
the bulk masses for these fields: $c \equiv M/k \simeq 1/2$ for the 
third generation and $c > 1/2$ for the first two generations (the size 
of the 4D Yukawa coupling is given by $\approx \sqrt{\ln(L_1/L_0)}\,g$, 
$\approx g$, and $\approx |c|\sqrt{\ln(L_1/L_0)}\,g\,(L_0/L_1)^{(|c|-1/2)}$ 
for $|c| < 1/2$, $=1/2$, and $>1/2$, respectively, where $g$ is the 4D 
gauge coupling). An additional ingredient to the flat space case is that 
some components of the bulk multiplets become exponentially light for 
$c > 1/2$. These are the fields with the following boundary conditions: 
$\Phi(+,-)$ and $\Phi^c(-,+)$ (in the 4D superfield notation) and give 
unwanted vector-like matter with masses much lighter than the TeV scale. 
We can, however, make these fields heavy by introducing appropriate 
fields $\bar{\Phi}$ and $\bar{\Phi}^c$ on the Planck and TeV branes, 
respectively, and by coupling them to $\Phi$ and $\Phi^c$ through 
the brane mass terms $\delta(z-L_0)[\bar{\Phi}\Phi]_{\theta^2}$ and 
$\delta(z-L_1)[\bar{\Phi}^c\Phi^c]_{\theta^2}$. [The corresponding 
terms in a non-supersymmetric theory are the brane fermion masses.] 
Proton decay is potentially dangerous in this theory, because 
SU(6)/(SU(3)$_C \times$SU(2)$_L \times$U(1)$_Y$) gauge fields 
have masses of order TeV, which could mediate rapid proton decay.
However, the structure of the theory allows us to impose the baryon 
number: ${\cal D}(1)$, ${\cal U}(1)$, ${\cal E}(0)$, ${\cal N}(0)$ 
with appropriate charges for the brane fields. Therefore, we can make 
proton absolutely stable. An important difference with respect to 
the flat space model is that, in the absence of brane-localized gauge 
kinetic terms, the gauge couplings in the present model should unify 
into SU(5)$\times$U(1)$_X$ at the TeV scale. As was explained before, 
this unwanted prediction is avoided if we introduce TeV-brane localized 
gauge kinetic terms such that the low-energy gauge coupling values 
are correctly reproduced,\footnote{
Large brane kinetic terms on the TeV brane may also 
help to reduce constraints from precision electroweak 
measurements~\cite{Davoudiasl:2002ua}.}
although it implies a loss of any quantitative prediction about 
gauge coupling unification. Small neutrino masses are obtained 
either by exponential suppressions of the neutrino Yukawa couplings 
or by the seesaw mechanism operated on the Planck brane in the case 
of large and small bulk right-handed neutrino masses, respectively.

The last issue toward a realistic theory of the PGB Higgs is the 
quartic coupling. If the theory is supersymmetric, as the one described 
above, the $\mathcal{O}(1)$ quartic coupling is generated through 
the supersymmetric gauge potential. Supersymmetric theories have 
two Higgs doublets at low energies, one of which is the PGB of 
the global symmetry.  The tree-level potential takes the form of 
$V_H \propto (|H_1|^2 - |H_2|^2)^2$, and the PGB Higgs corresponds 
to the direction $H_1 = H_2$. Supersymmetry can be broken either 
at the Planck brane~\cite{Gherghetta:2003wm} or at the TeV 
brane~\cite{Gherghetta:2000qt}. In the former case, the holographic 
theory is essentially a non-supersymmetric theory. Supersymmetry 
is a global invariance only of the CFT sector, and this partial 
supersymmetry is responsible for the generation of the tree-level 
quartic coupling in the Higgs potential without introducing a mass 
term. Having a non-zero Higgsino mass, however, will require some 
additional source of supersymmetry breaking. On the other hand, 
if supersymmetry is broken on the TeV brane in the 5D theory, 
the holographic theory is a locally supersymmetric theory with 
supersymmetry broken at the TeV scale by the CFT dynamics. The 
scale of supersymmetry breaking in this case should not be very high, 
as the tree-level Higgs quartic coupling becomes zero if the second 
Higgs boson, which is not a PGB, obtains a large supersymmetry 
breaking mass.

The quartic coupling in non-supersymmetric theories remains as 
a difficult issue. However, we can at least adopt the mechanism 
considered in the previous section with a little modification.
For example, in the case of an SU(3)$_L$ theory, we can introduce 
a bulk scalar field $\Phi$, transforming as a ${\bf 6}$ under 
SU(3)$_L$. By introducing a tadpole on the TeV brane for the 
SU(2)$_L$ singlet component of $\Phi$, we can generate the Higgs 
quartic coupling in essentially the same way as discussed in 
section~\ref{subsec:tree-higgs}. This possibility seems to indicate 
that we can have realistic theories in the non-supersymmetric 
case as well.

\section{Phenomenological Scales and Comparison with Pions 
in Large $N$ QCD}

To better understand the present theory of PGBs, it is instructive 
to look at the different physical scales of the model from the 
4D perspective. This will elucidate the PGB nature of the Higgs 
and its similarities with pions in QCD.

Let us consider the case in which the PGB is $A_5$ with the symmetry 
breaking pattern of Fig.~\ref{fig:breaking}. This is equivalent to 
the model of section~\ref{sec:higgsonb} if $v \gg \Lambda_\text{IR}$, 
since in this limit the SU(3)$_L$ breaking by $\Sigma$ reproduces the 
breaking by boundary conditions. The original 5D scales of the model 
are $\Lambda_{\rm IR}$, $1/g_5^2$, $k = L_0^{-1}$ and $L_1^{-1}$.
They can be related to 4D physical quantities in the following way:
\begin{equation}
  g_{\rho} \equiv \sqrt{g_5^2k}, \ \ \ \ 
  m_{\rho} \equiv \frac{\pi}{L_1}, \ \ \ \
  g^2 = \frac{g_{\rho}^2}{\ln(L_1/L_0)},
\label{eq:phs}
\end{equation}
where $g_{\rho}$ measures the strength of the KK coupling, $m_{\rho}$ 
is the mass splitting of the KK towers (approximately this is the 
first-KK mass), and $g$ is the 4D gauge coupling for the gauge bosons 
of $H$. Let us see how the different scales are related. Using naive 
dimensional analysis we can estimate $\Lambda_{\rm IR}$ (the scale at 
which the 5D gauge theory becomes strongly coupled for an observer 
on the TeV brane):
\begin{equation}
  \Lambda_{\rm IR} \approx \frac{24\pi^3}{g_5^2} \frac{L_0}{L_1}
  \approx \frac{24\pi^2 m_{\rho}}{g_{\rho}^2}.
\label{eq:lir}
\end{equation}
We can define a decay constant $f_{\pi}$ for the PGBs in our theory.
We follow the usual definition: $m_{W}^2 = g^2 f^2_{\pi}$, where 
$m_{W}$ is the mass that the gauge bosons $A_\mu^{\bar a}$ obtain 
from the strong dynamics. In the 5D picture this is the mass of the 
zero-mode gauge bosons with $(+,-)$ boundary conditions. One finds 
$m_{W}^2 = 2 /(L^2_1\ln(L_1/L_0))$, and therefore
\begin{equation}
  f_{\pi} = \frac{\sqrt{2}m_{\rho}}{\pi g_{\rho}}.
\label{eq:fpion}
\end{equation}
Using the AdS/CFT relation $g_{\rho} \sim 
1/\sqrt{N}$~\cite{Arkani-Hamed:2000ds}, we obtain 
$f_\pi \sim \sqrt{N}$ as expected in strongly coupled large $N$ 
theories~\cite{'tHooft:1973jz}. Notice that $\Lambda_{\rm IR}$ can 
be larger than the naive value of $4\pi f_\pi$. This is because the 
PGBs in our theory arise from higher dimensional gauge bosons, and 
the 5D gauge invariance improves the high energy behavior of the theory.
The smallest scale in the model is the mass of the PGBs. It appears 
at loop level, as that of charged pions in the massless quark limit.
We obtained in section~\ref{sec:A5higgs} that this is of order
\begin{equation}
  m_{\pi}^2 = m_{A_5}^2 
    \approx \frac{g^2}{16\pi^2}\, \frac{m^2_{\rho}}{\pi^2}.
\label{eq:pionmass}
\end{equation}
The value of the PGB mass from 5D shows the same dependence on 
$m_\rho$ as that of pions in QCD~\cite{Das:it}.

We then find that, in general, these theories have the following 
pattern of scales
\begin{equation}
  \Lambda_{\rm IR} > f_\pi > m_\rho > m_\pi.
\label{eq:pasc}
\end{equation}
In real QCD this pattern is not completely followed, since the pion 
decay constant is smaller than the $\rho$ mass. This could be due to 
the fact that in QCD we have $N=3$, which is not really a large number. 
In spite of this, other predictions of large $N$ QCD agree surprisingly 
well with the experimental data. Similarly, when the above 5D AdS model 
is used for the standard model, one realizes that the pattern of scales 
in Eq.~(\ref{eq:pasc}) is not really fulfilled. This is because 
$g_{\rho} \sim 4$ in order to reproduce the 4D gauge coupling values 
from $g^2 = g_{\rho}^2/\ln(L_1/L_0)$. Therefore, the theory of KK 
states (resonances) is very close to the non-perturbative limit
(in the large $N$ expansion). The pattern of scales that we obtain 
is similar to real QCD:
\begin{equation}
  \Lambda_{\rm IR} \,\simgt\, m_\rho > f_\pi > m_\pi.
\label{eq:pascr}
\end{equation}
We must emphasize, however, that the relation 
$g^2 = g_{\rho}^2/\ln(L_1/L_0)$ is subject to large logarithmic 
corrections and large dependence on brane-kinetic terms, so it is 
possible that $g_\rho$ takes smaller values than what are naively 
obtained from the 4D gauge coupling values, making the KK theory 
more perturbative. This is in fact also needed if we do not want to 
have large corrections to electroweak observables coming from virtual
KK states. These states couple to the Higgs (also a composite object
in these models) with a strength  $g_\rho$, and for the value 
$g_\rho\sim 4$ and $L_1\sim 1/{\rm TeV}$, they gives a too large 
deviation from the standard-model predictions.

\section{Conclusions}
\label{sec:concl}

We have presented a class of models where the standard-model Higgs 
appears as a composite PGB from a strongly coupled theory, similar 
to pions in QCD. We have used the AdS/CFT correspondence to describe 
the models in terms of the weakly coupled dual theory. The dual 
theory corresponds to a gauge theory in a slice of 5D AdS in which 
the bulk gauge symmetry is broken to the standard-model gauge group 
on both boundaries. This automatically delivers massless scalars 
(at tree level) on the TeV brane that we associate to the 
standard-model Higgs field.  A remnant global symmetry, under 
which the Higgs transforms non-linearly, protects the Higgs mass 
from large radiative corrections. The Higgs mass is generated at 
one-loop level through the explicitly breaking of the global symmetry 
due to the standard-model gauge interactions.  We have shown that 
this one-loop contribution is not quadratically divergent. In the 
5D AdS picture, this is because of the locality. The Higgs lives 
on the TeV brane away from the other scalar that breaks the bulk 
gauge symmetry (which is located on the Planck brane). Therefore, 
the Higgs can learn this breaking only by bulk fields that propagate 
from one brane to the other. This is a non-local effect and thus 
is finite. In the 4D CFT picture, the cancellation of quadratic 
divergences is understood in a different way. The Higgs is a composite 
state of CFT which decouples at high energies from the standard-model 
fields that are elementary states. We have calculated the effective 
potential of the Higgs from gauge loops and have shown that the Higgs 
squared-mass is finite and a loop factor smaller than the first 
resonance mass. This is a very appealing property, since it gives 
a rationale for the electroweak scale smaller than the new physics 
scale, as experiments seem to indicate.

If the breaking of the bulk gauge symmetry is due to boundary 
conditions, the massless scalar corresponds to the fifth component 
of the bulk gauge boson. Therefore, we find that Higgs-gauge 
unification in warped space is equivalent to a Higgs as a composite 
PGB. We have also discussed the similarities and differences of 
our PGBs with pions in QCD.

The models with the PGB Higgs generically suffer from the absence 
of a tree-level Higgs quartic coupling, needed to generate a physical 
Higgs mass larger than the experimental bound. We have presented 
a mechanism that can generate a quartic coupling without inducing 
a large quadratic term. This requires a specific assumption about 
the breaking of the global symmetry at high energies.

The models constructed here have an important phenomenological 
difference from little Higgs models. The global symmetry that 
protects the Higgs mass is a symmetry of the strong CFT sector
of the theory, but not a symmetry of the standard model.  Therefore, 
there is no partner of the standard-model fields to form a complete 
multiplet of the global symmetry. New electroweak-scale states appear 
as resonances that are in complete multiplets of the global group, 
similar to the situation in QCD. Detecting these resonances in future 
colliders will allow us to find the symmetries of the strong CFT, 
and tell us about the symmetry that protects the electroweak scale 
from potentially large radiative corrections.


\section*{Acknowledgments}

We would like to thank R.~Barbieri for useful discussions.
R.C. thanks P.~Creminelli, A.~Donini, B.~Gavela, E.~Trincherini 
and especially R.~Rattazzi for discussions.  Y.N. is grateful for 
valuable conversations with G.~Burdman and B.~Dobrescu, and 
A.P. thanks S.~Peris, M.~Quiros and E.~de~Rafael for discussions. 
The work of R.C. is supported by the EC under RTN contract 
HPRN-CT-2000-00148.


\section*{Appendix}

In this appendix we give the propagators for a bulk scalar, fermion and 
gauge fields in a slice of 5D AdS (see also Ref.~\cite{Gherghetta:2000kr}).  
We start by giving the scalar field propagator in the presence of 
general brane kinetic and mass terms.  The free action for a scalar 
field $\phi$ is given by
\begin{equation}
\begin{split}
  S =& \int\!d^4x \int_{L_0}^{L_1}\!\!dz 
    \Biggl\{ \sqrt{g} 
    \bigg[ g^{MN} \partial_M \phi^\dagger \partial_N \phi
    - M^2 \phi^\dagger \phi \biggr]
\\
  & + \delta(z-L_0) \sqrt{-g_{\rm ind}}
    \biggl[ z_0\, g_{\rm ind}^{\mu\nu}\, 
    \partial_\mu \phi^\dagger \partial_\nu \phi
    - m_0 \phi^\dagger \phi \biggr]
\\
  & + \delta(z-L_1) \sqrt{-g_{\rm ind}}
    \biggl[ z_1\, g_{\rm ind}^{\mu\nu}\, 
    \partial_\mu \phi^\dagger \partial_\nu \phi
    - m_1 \phi^\dagger \phi \biggr] \Biggr\}.
\label{eq:Ap-free-scalar}
\end{split}
\end{equation}
The propagator $\hat{G}$ is given as a solution of
\begin{equation}
  \left[ z^2 \partial_z^2 + z \partial_z 
  - \left( -p^2 z^2 + \alpha^2 \right) \right] 
  \hat{G}(z,z';p) = - \frac{z}{k} \delta(z-z').
\label{eq:Ap-prop-def}
\end{equation}
Here, $\alpha = \sqrt{4+M^2/k^2}$ and $\hat{G}$ represents the 
propagator for the rescaled field $\hat{\phi} \equiv (kz)^{-2}\phi$, 
which is related to the propagator $G$ for the unrescaled field, 
$\phi$, as $G = (kz)^2 (kz')^2 \hat{G}$.

If $\phi$ has boundary conditions $(+,+)$, the scalar propagator 
(for the rescaled field) is given by
\begin{equation}
\begin{split}
  \hat{G}(z,z';p) =& \frac{-L_0}{(X_I/X_K-Z_I/Z_K)} 
\\
  & \times \biggl( I_\alpha(|p|z_<)-\frac{Z_I}{Z_K} K_\alpha(|p|z_<) \biggr) 
     \biggl( I_\alpha(|p|z_>)-\frac{X_I}{X_K} K_\alpha(|p|z_>) \biggr),
\label{eq:Ap-prop}
\end{split}
\end{equation}
where $|p| \equiv \sqrt{-p^2}$ and $z_<$ ($z_>$) is the lesser 
(greater) of $z$ and $z'$; $I_\alpha(x)$ and $K_\alpha(x)$ are 
the modified Bessel functions. The coefficients $X_I$, $X_K$, 
$Z_I$ and $Z_K$ are given by
\begin{align}
\begin{split}
    X_I &= |p|L_1 I_{\alpha-1}(|p|L_1) 
        - \left( \alpha - s/2 - z_1 |p|^2 L_1^2 L_0^{-1} 
        - m_1L_0 \right) I_\alpha(|p|L_1), \\
    X_K &= -|p|L_1 K_{\alpha-1}(|p|L_1) 
        - \left( \alpha - s/2 - z_1 |p|^2 L_1^2 L_0^{-1} 
        - m_1L_0 \right) K_\alpha(|p|L_1),
\end{split}
\\[0.5cm]
\begin{split}
    Z_I &= |p|L_0 I_{\alpha-1}(|p|L_0) 
        - \left( \alpha - s/2 + z_0 |p|^2 L_0 
        + m_0 L_0 \right) I_\alpha(|p|L_0), \\
    Z_K &= -|p|L_0 K_{\alpha-1}(|p|L_0) 
        - \left( \alpha - s/2 + z_0 |p|^2 L_0 
        + m_0 L_0 \right) K_\alpha(|p|L_0),
\end{split}
\end{align}
where $s=4$.  The propagator for a field having the odd boundary 
condition at the Planck brane (TeV brane) is obtained by taking the 
limit $m_0 \rightarrow \infty$ ($m_1 \rightarrow \infty$).

Restricting the end points to the TeV brane, $z=z'=L_1$, and taking 
the zero-momentum limit, $p \rightarrow 0$, the scalar propagator of 
Eq.~(\ref{eq:Ap-prop}) becomes
\begin{equation}
\begin{split}
  \hat{G}(z=z'=L_1;&\, p\rightarrow 0) 
  = \frac{L_0}{2\alpha} \,
      \biggl\{ \frac{\alpha+2+m_1L_0}{\alpha-2-m_1L_0} 
      - \frac{\alpha+2-m_0L_0}{\alpha-2+m_0L_0}
      \left(\frac{L_0}{L_1}\right)^{2\alpha} \biggr\}^{-1}
\\
  & \times \biggl\{ 1 + \frac{\alpha+2-m_0L_0}{\alpha-2+m_0L_0}
      \left(\frac{L_0}{L_1}\right)^{2\alpha} \biggr\} 
    \biggl\{ 1 + \frac{\alpha+2+m_1L_0}{\alpha-2-m_1L_0} \biggr\}.
\label{eq:Ap-G}
\end{split}
\end{equation}
For $\alpha = 0$ ($M^2 = -4 k^2$), this is further simplified as
\begin{equation}
  \hat{G}(z=z'=L_1;\, p\rightarrow 0)
  = \frac{(1 + (m_0 L_0-2)\ln(L_1/L_0))}
    {m_0 + m_1 + L_0(m_0-2L_0^{-1})(m_1+2L_0^{-1})\ln(L_1/L_0)},
\label{eq:Ap-G-a0}
\end{equation}
giving the propagator used in section~\ref{subsec:tree-higgs} 
(Eq.~(\ref{eq:prop-TeV})).

The fermion propagators used in section~\ref{subsec:mass-comp} are 
given by
\begin{equation}
  S^{(\pm,\pm)}(z,z^\prime;p) 
  = -\left(k^2 zz^\prime\right)^{5/2} 
    \left[-\!\not\! p+\gamma^5 \left(\partial_z+\frac{1}{2z}\right) 
    + \frac{M_\Psi}{kz}\right] 
    \left(P_R\,\hat{G}_R^{(\pm,\pm)} + P_L\,\hat{G}_L^{(\mp,\mp)}\right),
\label{eq:Ap-propfirst}
\end{equation}
where $P_{L,R}=(1\mp\gamma^5)/2$ and $M_\Psi$ is the bulk 
mass of the fermion. The quantity $\hat{G}_R^{(+,+)}$ is 
given by Eq.~(\ref{eq:Ap-prop}) for $\alpha=|M_\Psi/k+1/2|$, $s=1$, 
$m_0=-M_\Psi$, $m_1=M_\Psi$ and $z_0=z_1=0$. The case of $\hat{G}_R$ 
with the odd boundary condition at the Planck brane (TeV brane) is 
reproduced by taking the limit $m_0 \to \infty$ ($m_1 \to \infty$). 
The expression for $\hat{G}_L$ is obtained from that of $\hat{G}_R$ 
by simply making the replacement $M_\Psi \to -M_\Psi$.

Finally, the rescaled gauge boson propagator is given by taking 
$\alpha =1$ and $s=2$ in Eq.~(\ref{eq:Ap-prop}). The parameters $m_0$ 
and $m_1$ then represent brane masses for the gauge boson induced by 
spontaneous symmetry breaking caused by brane Higgs fields. The case 
of boundary condition breaking at the Planck brane (TeV brane) is 
reproduced by taking the limit $m_0 \to \infty$ ($m_1 \to \infty$).


\end{document}